\documentclass[journal,onecolumn,11pt,draftclsnofoot]{IEEEtran}

\usepackage[a4paper,margin=2.5cm]{geometry}
\usepackage[T1]{fontenc}
\usepackage[utf8]{inputenc}
\usepackage{amsmath,amssymb,amsthm}
\usepackage{mathtools}
\usepackage{mathrsfs}
\usepackage{hyperref}
\usepackage{pifont}
\usepackage{framed}
\usepackage{array}
\usepackage{booktabs}
\usepackage{longtable}
\usepackage{float}

\usepackage{algorithm}
\usepackage{algpseudocode}
\usepackage{xcolor}
\usepackage{listings}
\usepackage{pgfplots}
\pgfplotsset{compat=1.17}
\usepackage{enumitem}

\hypersetup{colorlinks=true, linkcolor=blue, citecolor=blue, urlcolor=blue}

\theoremstyle{definition}
\newtheorem{definition}{Definition}[section]
\newtheorem{counterexample}[definition]{Counterexample}
\theoremstyle{plain}
\newtheorem{proposition}[definition]{Proposition}
\newtheorem{lemma}[definition]{Lemma}
\newtheorem{theorem}[definition]{Theorem}
\newtheorem{corollary}[definition]{Corollary}

\theoremstyle{remark}
\newtheorem{remark}[definition]{Remark}

\newcommand{\Sig}{\Sigma}
\newcommand{\Sigstar}{\Sigma^*}
\newcommand{\Sdict}{S}
\newcommand{\sub}[1]{\operatorname{Fact}(#1)}
\newcommand{\Ssub}{\operatorname{Fact}(\Sdict)}

\newcommand{\CS}[2][\Sdict]{C_{#1}(#2)}
\newcommand{\CadaptSym}[1][\Sdict]{C_{#1}^{\mathrm{adapt}}}
\newcommand{\Cadapt}[2][\Sdict]{C_{#1}^{\mathrm{adapt}}(#2)}
\newcommand{\Cctx}[3][\Sdict]{C_{#1}^{\mathrm{adapt}}(#2 \mid #3)}

\newcommand{\len}[1]{|#1|}

\title{On the Adaptive Covering Complexity of Strings: Unifying Relative and Self-Referential Compression}
\author{Pierre-François Marteau\\IRISA, Université de Bretagne Sud\\pierre-francois.marteau@univ-ubs.fr}

\begin{document}
	\maketitle

\begin{abstract}
	We introduce and study the adaptive covering complexity $C_S^{\mathrm{adapt}}(s)$, defined as the minimum number of contiguous factor blocks required to partition a string $s$ when each block is either a factor of a fixed external dictionary $S$ or a factor of the target's already decoded prefix. This hybrid admissibility model formalizes the structural interaction between static dictionary matching and dynamic non-overlapping self-referencing. 
	
	Our main theoretical result is the exact logarithmic law $1+\lceil\log_2 C_S(s)\rceil \le C_S^{\mathrm{adapt}}(s) \le C_S(s)$, which is tight and proves that dynamic self-reference can reduce phrase count by at most an exponential factor over static dictionary covering alone. We establish structural equivalences showing that the measure bridges word-segmented Relative Lempel--Ziv and non-overlapping LZ77, and we prove exact preset-history LZ77 sandwich bounds. Furthermore, we establish the optimality of the longest-match greedy parse via an exchange lemma and present an optimal implementation based on dual suffix automata running in $O(M+n)$ time and space for a constant alphabet. These analytical results are complemented by counting bounds, stationary-ergodic asymptotic invariance, a dictionary-selection NP-hardness result, and a concise empirical illustration.
\end{abstract}
	

\section{Introduction}
\label{sec:intro}

A central theme in source coding and string parsing is the exploitation of sequence redundancy through dictionary-based mechanisms. When encoding finite strings, dictionary models typically draw references from two distinct contexts: \emph{static external references}, which provide global background knowledge independent of the target sequence (as in Relative Lempel--Ziv or preset-dictionary compression), and \emph{dynamic internal references}, which exploit local self-similarity from the already decoded prefix (as in classical LZ77). While both paradigms are foundational to practical compression, their theoretical interaction—specifically, the structural trade-off between external static memory and growing internal memory—has been less studied as a single phrase-count measure within combinatorial parsing theory, although closely related greedy-parsing results exist (Section~\ref{sec:related}).

In this paper, building on the covering measure of~\cite{marteau2019}, which we restrict to contiguous factors and extend with self-reference, we introduce and study the \emph{adaptive covering complexity}, denoted $C_S^{\mathrm{adapt}}(s)$. Given a fixed external dictionary $S$, $C_S^{\mathrm{adapt}}(s)$ measures the minimum number of contiguous factor blocks required to partition a target string $s$ under a hybrid admissibility rule: each block must be either a factor of a word in $S$ or a factor of the target's already decoded prefix. This combinatorial measure provides a clean theoretical bridge, isolating the phrase-count consequences of combining global static dictionaries with non-overlapping local self-referencing.

It is crucial to distinguish this theoretical framework from practical compression implementations (such as zlib, Zstandard, or Brotli). While practical systems optimize a bit-level description cost using dynamic programming, sliding windows, and entropy coding, $C_S^{\mathrm{adapt}}(s)$ is a purely combinatorial phrase-count measure. Furthermore, while practical preset-dictionary LZ77 algorithms concatenate external history with target text into a single search buffer, $C_S^{\mathrm{adapt}}$ keeps static external dictionary matching and dynamic internal self-referencing structurally distinct. This formal separation enables us to establish exact mathematical laws and structural bounds that cannot be derived directly from standard LZ77 parsing rules or bit-cost heuristics.

Our investigation addresses four foundational questions regarding this dual-source model:
\begin{enumerate}[leftmargin=*]
	\item How can adaptive covering be formally defined to unify static dictionary covering and dynamic self-referencing within a single non-overlapping admissibility framework?
	\item What exact structural and quantitative laws govern adaptive covering, and how do they bound the gain of dynamic self-reference relative to static dictionary covering $C_S(s)$?
	\item How does this combinatorial complexity relate to classical Relative Lempel--Ziv, non-overlapping LZ77, and preset-history LZ77 parsing?
	\item Can the optimal greedy adaptive parse be computed efficiently, and what is its computational complexity?
\end{enumerate}

We emphasize that minimum phrase count is a combinatorial measure distinct from bit-level description cost. To complement our theoretical analysis, we provide a compact empirical section summarizing a dual-source bit-cost codec featuring Minimum Description Length (MDL) dictionary selection, while deferring full implementation details, ablations, and supplementary structural results to the appendices. None of this appendix material is used in the proofs of Sections~\ref{sec:theory} and~\ref{sec:algorithm}, which contain all the main results: Appendices~\ref{app:structural} and~\ref{app:codec}, and Section~\ref{sec:main-experiments} that summarizes the latter, can be removed, for instance in a shortened version, without affecting the main theorems.

\paragraph*{Main Contributions.}
\begin{enumerate}[leftmargin=*]
	\item \textbf{Formal Framework:} We formulate the adaptive covering complexity $C_S^{\mathrm{adapt}}(s)$, establishing a rigorous combinatorial bridge between external dictionary matching and internal self-referencing.
	\item \textbf{Exact Logarithmic Law:} We prove the tight bounds $1+\lceil\log_2 C_S(s)\rceil \le C_S^{\mathrm{adapt}}(s) \le C_S(s)$ (Theorem~\ref{thm:exp}), demonstrating that dynamic self-reference can reduce phrase count by at most an exponential factor. The lower bound generalizes to an arbitrary dictionary the classical doubling argument for non-self-referential LZ parsing, in which each phrase at most doubles the decoded text; for $S=\emptyset$ it reduces to the fact that such a parse of a string of length $n$ has at least $1+\lceil\log_2 n\rceil$ phrases. We also derive a sharp repetition bound $C_S^{\mathrm{adapt}}(w^n) \le C_S^{\mathrm{adapt}}(w) + \lceil\log_2 n\rceil$.
	\item \textbf{Parsing Equivalences and Sandwich Theorems:} Under the non-overlapping factor convention, we prove that $C_\emptyset^{\mathrm{adapt}}(s)$ coincides with non-overlapping LZ77 phrase count $z_{\mathrm{no}}(s)$, and that $C_S^{\mathrm{adapt}}(s)$ is tightly sandwiched with preset-history LZ77 up to an exact factor of 2 (Theorem~\ref{thm:sandwich}).
	\item \textbf{Optimal Greedy Parsing and Linear-Time Algorithm:} We establish the optimality of greedy parsing via an exchange lemma (Theorem~\ref{thm:greedy-choice}) and present a dual suffix-automaton implementation running in $O(M+n)$ time and space for constant alphabets (Theorem~\ref{thm:complexity}).
	\item \textbf{Supplementary Analysis and Empirical Illustration:} We provide counting bounds, stationary-ergodic asymptotic invariance, and a dictionary-selection NP-hardness reduction in the appendices, alongside a compact empirical evaluation illustrating practical dual-source encoding. These results are not needed for the main theorems and may be omitted if required.
\end{enumerate}

\section{Related Work}
\label{sec:related}

Adaptive covering sits at the intersection of relative dictionary compression, Lempel--Ziv parsing, combinatorics on words, and data structures for factor matching. The closest literature is not limited to practical preset-dictionary codecs: in particular, greedy phrase-count optimality and asymptotic LZ parsing already have substantial theoretical histories. We therefore distinguish established results from the specific combination of admissible sources studied here.

\paragraph*{Relative compression and preset dictionaries.}
Relative Lempel--Ziv (RLZ) parses a target against a fixed reference and has been studied for collections of related texts, including web collections and genomes~\cite{kuruppu2010relative,hoobin2011}, in the spirit of the cross-parsing of Ziv and Merhav~\cite{zivmerhav1993}. Practical formats also allow a preset dictionary to initialize the history available to an LZ encoder; zlib, Zstandard, and Brotli are prominent examples~\cite{deutsch1996zlib,collet2016zstandard,alakuijala2018brotli}. Dictionary construction is itself a design problem, and methods for constructing effective relative-LZ dictionaries and trained buffers address that engineering objective~\cite{liao2016}. These systems establish that external and previously decoded material can both be useful in compression. They do not, by themselves, imply the structural bounds proved here: practical codecs optimize a bitstream under format-specific match, window, and coding constraints, whereas $C_S^{\mathrm{adapt}}$ counts admissible contiguous factors.

A modeling distinction matters when comparing with RLZ. In this paper, $S$ is a set of dictionary words and $\operatorname{Fact}(S)$ is the union of factors contained in individual words. A factor that crosses the boundary between two reference words is therefore not automatically admissible. This convention is narrower than treating a concatenation of references as one reference string, and the comparison with RLZ below is made under the stated word-segmented convention.

\paragraph*{Greedy parsing: phrase count versus bit cost.}
The foundational Lempel--Ziv schemes describe strings by references to earlier material~\cite{zivlempel1977,zivlempel1978}; Storer and Szymanski analyze textual-substitution schemes and the role of overlap~\cite{storer1982data}. For phrase count, longest-match greedy parsing is known to be optimal in classical LZ77 settings, and work on suffix-closed dynamic dictionaries establishes conditions under which this property persists~\cite{greedyparsing2014}. Cohn and Khazan studied parsing with prefix and suffix dictionaries~\cite{cohn1996parsing}, while Langiu surveys the broader distinction between minimizing phrase count and minimizing encoded bit length~\cite{langiu2013}. Kosolobov and Shur compare overlapping and non-overlapping LZ77 variants and show that their phrase counts can differ by a tight logarithmic factor~\cite{kosolobov2019comparison}. We therefore do not claim a new general principle of greedy parsing. The self-contained exchange proof instead verifies optimality for the precise hybrid admissibility rule used here: a phrase must be a factor either of a fixed external dictionary word or of the already decoded prefix, with self-copies required to lie wholly in available material and not overlap the target phrase. This distinction matters because phrase-count optimality does not imply bit-cost optimality.

\paragraph*{Asymptotic LZ theory.}
The normalized phrase count of Lempel--Ziv parsing has long been studied for stationary ergodic sources. Ornstein and Weiss relate typical LZ phrase lengths to source entropy and provide a central asymptotic foundation for this line of work~\cite{ornsteinweiss1993}. These results explain why the leading-order phrase-count behavior is governed by the source rather than by a fixed finite initial context. Our asymptotic result is related but has a specific scope: it compares the adaptive model with and without a fixed finite dictionary under the assumptions stated in Theorem~\ref{thm:asymptotic_invariance}. It asserts equality of asymptotic normalized phrase counts, not equality of finite-length parses, encoded bit rates, or practical compression ratios.

\paragraph*{Factor automata and algorithmic matching.}
Suffix automata and factor automata represent the set of substrings of a word or a collection of words and support efficient factor matching. Blumer et al. give the classical linear-size automaton for the subwords of a text~\cite{blumer1985}; Mohri, Moreno, and Weinstein extend the construction and analysis to sets of strings~\cite{mohri2009}. These data structures are established tools, not a contribution claimed in isolation here. Our algorithmic contribution is to combine indexes for the external dictionary and the growing decoded prefix with the greedy decision rule, and to analyze the total cost of the resulting adaptive parse as $O(M+n)$ under the implementation assumptions specified in Section~\ref{sec:algorithm}. The linear bound depends on the matching procedure and amortization argument given there, rather than following merely from the existence of a suffix automaton.

\paragraph*{Covering complexity, grammar compression, and dictionary selection.}
Marteau introduced the covering measure from which the static quantity $C_S(s)$, restricted here to contiguous factors, is derived~\cite{marteau2019}; it serves as the static point of comparison. Grammar-based compression and smallest-grammar complexity provide a related but different perspective: they allow a reusable generative structure, rather than requiring each phrase to be a contiguous factor of an external word or the already decoded prefix~\cite{rytter2003,charikar2005}. The NP-hardness result for choosing a dictionary is connected to the Shortest Common Superstring problem, whose complexity has been studied directly in the string-algorithms literature~\cite{superstring1994}. This reduction establishes hardness of the optimization problem under our dictionary model; it does not establish NP-completeness unless membership in NP is also proved for the chosen decision formulation.

Finally, the codec's dictionary-selection objective follows the Minimum Description Length principle and its close relative, the minimum-description approach to statistical modeling~\cite{grunwald2007minimum,rissanen1978modeling}. This is conceptually distinct from the combinatorial measure: MDL accounts for the cost of describing the dictionary as well as the encoded records, while $C_S^{\mathrm{adapt}}$ counts phrases without assigning them bit costs. The experiments therefore test one concrete bit-cost realization of the model and should not be interpreted as proving that phrase-count optimality implies bit-optimal compression.

\paragraph*{Positioning and scope.}
The contribution is the analysis of one explicit hybrid admissibility rule, rather than the first use of preset dictionaries, the first greedy LZ parse, or the first suffix-automaton index. For this rule, we prove the logarithmic relation between static and adaptive covering, establish the stated relations with selected RLZ and LZ77 phrase-count measures, and give an optimal greedy algorithm with a linear-time implementation under the paper's assumptions. The counting and source-asymptotic results are complementary structural statements; they do not constitute bit-cost bounds. This separation is important because the codec's practical performance depends on dictionary/header overhead and on the actual coding costs of literals and pointers.

\section{Theoretical Foundations of \texorpdfstring{$\CadaptSym$}{C\_S\^{}adapt}}
	\label{sec:theory}
	
	
	\subsection{Definitions and Modeling}
	
	Let $\Sigma$ be a non-empty finite alphabet and $\Sigma^*$ the set of finite strings over $\Sigma$. Unless explicitly stated otherwise, no lower bound on $|\Sigma|$ is needed for the finite-string results. Let $S\subseteq\Sigma^*$ be a finite external dictionary. For a string $x$, let $\operatorname{Fact}(x)$ denote the set of its non-empty contiguous factors, and for a set $X$ let
	\[
	\operatorname{Fact}(X) = \left(\bigcup_{x\in X}\operatorname{Fact}(x)\right)\cup\Sigma.
	\]
	Thus literals are always admissible. Throughout the paper, \emph{reference} means an occurrence entirely contained in material already available to the decoder; overlapping/self-extending copies are not part of the model.
	
	\begin{definition}[Covering]
		A covering of a non-empty $s\in\Sigma^*$ with respect to $S$ is a partition $s=p_1\cdots p_k$ such that $p_i\in\operatorname{Fact}(S)$ for every $i$. We write $C_S(s)$ for the minimum possible $k$, with $C_S(\epsilon)=0$.
	\end{definition}
	
	\begin{definition}[Adaptive covering]
		For $s=s_1\cdots s_n$ and a position $i\in\{1,\ldots,n\}$, define the available source set
		\[
		A_i(S,s)=\operatorname{Fact}(S)\cup\operatorname{Fact}(s[1..i-1]).
		\]
		An adaptive covering is a partition $s=p_1\cdots p_k$ such that, if $p_1\cdots p_{r-1}=s[1..i-1]$, then $p_r\in A_i(S,s)$. The adaptive covering complexity is
		\[
		C_S^{\mathrm{adapt}}(s)=\min\{k:s\text{ admits such a covering with }k\text{ blocks}\},\qquad C_S^{\mathrm{adapt}}(\epsilon)=0.
		\]
		Admissibility depends on the decoded prefix as a string, not on its phrase boundaries.
	\end{definition}
	
	\begin{definition}[Greedy adaptive parse]\label{def:greedy-parse}
		At position $i$, let
		\[
		L_i(S,s)=\max\{\ell\ge1:s[i..i+\ell-1]\in A_i(S,s)\},
		\]
		with $\ell\le n-i+1$. The greedy adaptive parse takes $g_i=s[i..i+L_i(S,s)-1]$ and advances to $i+L_i(S,s)$. Its number of blocks is denoted $G_S(s)$.
	\end{definition}
	
	\begin{remark}[Exact modeling choice]
		The original covering formulation associated with $\operatorname{sub}(S)$ uses the word ``subsequence''. Here we explicitly restrict the model to \emph{contiguous factors}, because the proposed algorithm and the LZ connection are factor-based. This is a deliberate restriction, not a silent reinterpretation of a cited definition.
	\end{remark}
	
	\begin{remark}[Relation to relative and self-referential parsing]
		With only $\operatorname{Fact}(S)$ available, the model is a static dictionary covering. Adding $\operatorname{Fact}(s[1..i-1])$ makes the source set grow with the decoded target prefix. Fixed external references are available from the first position; self-references become available as decoding proceeds.
	\end{remark}
	
	\subsection{Properties of the non-adaptive scheme \texorpdfstring{$\CS{\cdot}$}{C\_S}}
	
	\begin{proposition}[Bounds]\label{prop:bounds}
		For every non-empty $s \in \Sigstar$ and every $\Sdict \subset \Sigstar$, $1 \le \CS{s} \le \len{s}$. Moreover $\CS{s} = 1 \iff s \in \Ssub$.
		The same bounds hold for $\Cadapt{s}$.
	\end{proposition}
	
	\begin{proof}
		The lower bound is immediate. The upper bound follows from the covering by single symbols: since $\Sig \subset \Ssub$, the partition
		into $\len{s}$ blocks of length $1$ is always valid. If $s \in \Ssub$, $c=\{s\}$ is a valid covering, so $\CS{s}\le 1$. Conversely, if
		$\CS{s}=1$ the unique block $p_1=s$ lies in $\Ssub$. The adaptive case is identical.
	\end{proof}
	
	\begin{proposition}[Monotonicity in the dictionary]\label{prop:monotonicity}
		If $S_1 \subseteq S_2$, then for every $s \in \Sigstar$, $\CS[S_2]{s} \le \CS[S_1]{s}$ and $\Cadapt[S_2]{s}\le\Cadapt[S_1]{s}$. In particular, in phrase count the dictionary $S_1\cup S_2$ is at least as good as either $S_1$ or $S_2$, for both measures.
	\end{proposition}
	
	\begin{proof}
		$S_1\subseteq S_2$ implies $\sub{(S_1\cup X)}\subseteq\sub{(S_2\cup X)}$ for any $X$, so every covering
		(adaptive or not) with respect to $S_1$ is also one with respect to $S_2$.
	\end{proof}
	
	\begin{lemma}[Monotonicity by contiguous factor]\label{lem:factor-monotone}
		If $s_1$ is a contiguous factor of $s_2$, then for every $\Sdict \subset \Sigstar$, $\CS{s_1} \le \CS{s_2}$.
	\end{lemma}
	
	\begin{proof}
		Let $c_{\Sdict}^*(s_2) = (p_1,\dots,p_k)$ be an optimal covering of $s_2$. Restricting each block to the index range occupied by $s_1$
		inside $s_2$ yields a partition of $s_1$ into at most $k$ blocks (those that intersect the range). Each restricted block is a
		contiguous factor of an element of $\Ssub$, hence still belongs to $\Ssub$.
	\end{proof}
	
	\begin{remark}
		Lemma~\ref{lem:factor-monotone} is specific to the non-adaptive scheme. For $\CadaptSym$ the restriction argument breaks down, since
		cutting away the beginning of $s_2$ removes part of the prefix that later blocks refer to. The property is in fact false: for
		$\Sdict=\emptyset$, $s_2=\texttt{aabbaaaab}$ has $\Cadapt[\emptyset]{s_2}=6$ whereas its factor $\texttt{abbaaaab}$ has
		$\Cadapt[\emptyset]{\texttt{abbaaaab}}=7$.
	\end{remark}
	
	\begin{proposition}[Exact bracketing of concatenation]\label{prop:exact-concat}
		For all non-empty $s_1, s_2 \in \Sigstar$ and every $\Sdict \subset \Sigstar$,
		\[ \CS{s_1}+\CS{s_2} - 1 \;\le\; \CS{s_1\cdot s_2} \;\le\; \CS{s_1}+\CS{s_2}. \]
	\end{proposition}
	
	\begin{proof}
		The upper bound follows by concatenating two optimal coverings. For the lower bound, let $c = (q_1, \dots, q_m)$ be an $S$-optimal covering of $s_1 \cdot s_2$. The boundary separating $s_1$ and $s_2$ in $s_1 \cdot s_2$ (i.e., the exact position immediately following the last character of $s_1$) either falls between two blocks (so $C_S(s_1) + C_S(s_2) \le m$) or cuts a block $q_j = u \cdot v$ with $u, v \in Fact(S)$ (so $C_S(s_1) \le j$ and $C_S(s_2) \le m-j+1$, i.e., $C_S(s_1) + C_S(s_2) \le m+1$).
	\end{proof}
	
	\begin{corollary}[Powers]\label{cor:powers-exact}
		For every non-empty $s \in \Sigstar$, every $\Sdict \subset \Sigstar$ and every $n \ge 1$,
		$n\, \CS{s} - (n-1) \le \CS{s^n} \le n\, \CS{s}$.
	\end{corollary}

	\begin{proof}
		For $n=1$ both bounds are equalities. If they hold for $n-1$, Proposition~\ref{prop:exact-concat} applied to $s^{n-1}$ and $s$ gives $\CS{s^n}\le\CS{s^{n-1}}+\CS{s}\le n\,\CS{s}$ and $\CS{s^n}\ge\CS{s^{n-1}}+\CS{s}-1\ge n\,\CS{s}-(n-1)$.
	\end{proof}
	
	\begin{counterexample}[Static covering does not support adaptive prefix reuse]\label{cex:no-chain}
		Let $\Sdict = \{\texttt{a}\}$ and $s_1 = s_2 = \texttt{bc}$. Then $\CS{s_1}=2$ and $\Cadapt[\Sdict\cup\{s_1\}]{s_2}=1$, while $\CS{s_1s_2}=\CS{\texttt{bcbc}}=4$. Thus allowing the first string as an external word for the second changes the admissible source set; it cannot be modeled by simply concatenating the two static covering costs.
	\end{counterexample}
	
	\subsection{Properties of the adaptive variant \texorpdfstring{$\CadaptSym$}{C\_S\^{}adapt}}
	\label{sec:adaptive}
	
	Unlike the non-adaptive scheme $\CS{\cdot}$, the $\CadaptSym$ variant satisfies properties that are essential for compression. To state the chain rule precisely, we define the cost of covering $s_2$ \emph{in the context of} $s_1$, i.e.\ the cost of the second half of $s_1s_2$ when the first half is already available as prefix.
	
	\begin{definition}[Contextual adaptive covering]
		For non-empty strings $s_1,s_2\in\Sigma^*$, let
		$C_S^{\mathrm{adapt}}(s_2\mid s_1)$ be the minimum number of blocks in a
		factorization $s_2=q_1\cdots q_\ell$ such that, for every $j$,
		\[
		q_j\in
		\operatorname{Fact}(S)
		\cup
		\operatorname{Fact}(s_1q_1\cdots q_{j-1}).
		\]
	\end{definition}
	
	\begin{proposition}[Adaptive chain rule]
		\label{prop:chain-adapt}
		For all non-empty $s_1,s_2\in\Sigma^*$ and every finite dictionary
		$S\subseteq\Sigma^*$,
		\[
		C_S^{\mathrm{adapt}}(s_1)
		+
		C_S^{\mathrm{adapt}}(s_2\mid s_1)
		-1
		\;\le\;
		C_S^{\mathrm{adapt}}(s_1s_2)
		\;\le\;
		C_S^{\mathrm{adapt}}(s_1)
		+
		C_S^{\mathrm{adapt}}(s_2\mid s_1).
		\]
		Moreover,
		\[
		C_S^{\mathrm{adapt}}(s_2\mid s_1)
		\le
		C_{S\cup\{s_1\}}^{\mathrm{adapt}}(s_2),
		\]
		and consequently
		\[
		C_S^{\mathrm{adapt}}(s_1s_2)
		\le
		C_S^{\mathrm{adapt}}(s_1)
		+
		C_{S\cup\{s_1\}}^{\mathrm{adapt}}(s_2).
		\]
	\end{proposition}
	
	\begin{proof}
		We first prove the upper bound. Let
		\[
		s_1=p_1\cdots p_r
		\]
		be an optimal adaptive covering of $s_1$, so that
		\[
		r=C_S^{\mathrm{adapt}}(s_1),
		\]
		and let
		\[
		s_2=q_1\cdots q_\ell
		\]
		be an optimal covering of $s_2$ in the context of $s_1$, so that
		\[
		\ell=C_S^{\mathrm{adapt}}(s_2\mid s_1).
		\]
		By definition of contextual admissibility, each $q_j$ is a factor of an
		element of $S$ or of
		\[
		s_1q_1\cdots q_{j-1}.
		\]
		Therefore the concatenation
		\[
		p_1\cdots p_rq_1\cdots q_\ell
		\]
		is an adaptive covering of $s_1s_2$. Hence
		\[
		C_S^{\mathrm{adapt}}(s_1s_2)
		\le r+\ell,
		\]
		which gives
		\[
		C_S^{\mathrm{adapt}}(s_1s_2)
		\le
		C_S^{\mathrm{adapt}}(s_1)
		+
		C_S^{\mathrm{adapt}}(s_2\mid s_1).
		\]
		
		For the lower bound, let
		\[
		s_1s_2=b_1\cdots b_m
		\]
		be an optimal adaptive covering, with
		\[
		m=C_S^{\mathrm{adapt}}(s_1s_2).
		\]
		Consider the boundary between $s_1$ and $s_2$.
		
		If the boundary falls between two blocks, say after $b_j$, then
		\[
		s_1=b_1\cdots b_j
		\qquad\text{and}\qquad
		s_2=b_{j+1}\cdots b_m.
		\]
		The first part is an adaptive covering of $s_1$, while the second part
		is a valid contextual covering of $s_2$ given $s_1$. Consequently,
		\[
		C_S^{\mathrm{adapt}}(s_1)
		+
		C_S^{\mathrm{adapt}}(s_2\mid s_1)
		\le
		j+(m-j)=m.
		\]
		
		Otherwise, the boundary cuts a unique block, say
		\[
		b_j=uv,
		\]
		where $u$ is a non-empty suffix of $s_1$ and $v$ is a non-empty prefix
		of $s_2$. Since $b_j$ is admissible at its position, every contiguous
		factor of $b_j$ is admissible from the same source. Hence
		\[
		b_1,\ldots,b_{j-1},u
		\]
		is an adaptive covering of $s_1$ with $j$ blocks.
		
		For the second part, consider
		\[
		v,b_{j+1},\ldots,b_m.
		\]
		At the beginning of this contextual covering, the whole string $s_1$
		is available. Since $v$ is a factor of the admissible block $b_j$, it
		is also admissible. After decoding $v$, the available context is
		exactly $s_1v$, and thereafter the original blocks
		$b_{j+1},\ldots,b_m$ remain admissible because their original source
		occurrences were already available before their corresponding target
		blocks were decoded. Thus
		\[
		v,b_{j+1},\ldots,b_m
		\]
		is a contextual covering of $s_2$ with $m-j+1$ blocks.
		
		Therefore
		\[
		C_S^{\mathrm{adapt}}(s_1)
		+
		C_S^{\mathrm{adapt}}(s_2\mid s_1)
		\le
		j+(m-j+1)=m+1,
		\]
		which is equivalent to
		\[
		C_S^{\mathrm{adapt}}(s_1)
		+
		C_S^{\mathrm{adapt}}(s_2\mid s_1)
		-1
		\le
		C_S^{\mathrm{adapt}}(s_1s_2).
		\]
		
		It remains to compare contextual covering with a dictionary containing
		$s_1$. Let
		\[
		s_2=q_1\cdots q_\ell
		\]
		be an optimal adaptive covering with respect to $S\cup\{s_1\}$.
		For every $q_j$, its source is either a factor of an element of $S$, a
		factor of $s_1$, or a factor of the already decoded prefix
		$q_1\cdots q_{j-1}$. In all three cases the same source is available in
		the contextual model with initial context $s_1$. Hence the same covering
		is valid contextually, and
		\[
		C_S^{\mathrm{adapt}}(s_2\mid s_1)
		\le
		C_{S\cup\{s_1\}}^{\mathrm{adapt}}(s_2).
		\]
		Combining this inequality with the upper bound already proved yields
		\[
		C_S^{\mathrm{adapt}}(s_1s_2)
		\le
		C_S^{\mathrm{adapt}}(s_1)
		+
		C_{S\cup\{s_1\}}^{\mathrm{adapt}}(s_2).
		\]
	\end{proof}
	
	\begin{remark}
		The lower bound cannot be stated with $\Cadapt[\Sdict\cup\{s_1\}]{s_2}$ in place of $\Cctx{s_2}{s_1}$: this dictionary
		does not contain factors straddling the boundary between $s_1$ and $s_2$. For instance, with $\Sdict=\{\texttt{babb}\}$, $s_1=\texttt{aa}$
		and $s_2=\texttt{bbaabbb}$ one finds $\Cadapt{s_1}=2$, $\Cadapt[\Sdict\cup\{s_1\}]{s_2}=4$, but
		$\Cadapt{s_1s_2}=4<2+4-1$ (cover $\texttt{aabbaabbb}$ as $\texttt{a}\,|\,\texttt{abb}\,|\,\texttt{aabb}\,|\,\texttt{b}$).
	\end{remark}
	
	\begin{proposition}[Repetition bound]\label{prop:idem-adapt}
		For non-empty $w \in \Sigstar$ and $n \ge 1$,
		\[ \Cadapt{ww} \le \Cadapt{w} + 1 \qquad\text{and}\qquad \Cadapt{w^n} \le \Cadapt{w} + \lceil\log_2 n\rceil. \]
		The second bound is tight: $\Cadapt[\emptyset]{a^n}=1+\lceil\log_2 n\rceil$ for every letter $a$.
	\end{proposition}
	
	\begin{proof}
		\emph{Two copies.} After an optimal adaptive covering of the first occurrence of $w$, the covered prefix is exactly $w$; the second occurrence is a factor of it and forms one additional block.
		
		\emph{Doubling.} Suppose $w^m$ with $1\le m<n$ has been covered. The next block $w^{\min(m,n-m)}$ is a factor of the prefix $w^m$, hence admissible, and afterwards $w^{\min(2m,n)}$ is covered. Starting from $m=1$, which costs $\Cadapt{w}$, reaching $m=n$ takes $\lceil\log_2 n\rceil$ further blocks.
		
		\emph{Tightness.} For $S=\emptyset$ and $w=a$, the upper bound is attained. Indeed, the first phrase has length one, and each subsequent non-overlapping phrase has length at most the length of the already decoded prefix. After $k$ phrases, at most $2^{k-1}$ symbols can therefore have been covered. Thus any parse of $a^n$ needs at least $1+\lceil\log_2 n\rceil$ phrases, matching the construction above.
	\end{proof}
	
	Periodic structure is thus covered with logarithmically many blocks, where a static dictionary needs linearly many (Counterexample~\ref{cex:no-chain} and Corollary~\ref{cor:floor}).
	
	\subsection{Adaptive versus static covering: an exact logarithmic law}
	\label{sec:logarithmic-law}
	
	The following theorem quantifies how much the decoded prefix can save with respect to the dictionary alone.
	
	\begin{theorem}[Logarithmic law]\label{thm:exp}
		For every finite $S\subset\Sigstar$ and every non-empty $s\in\Sigstar$,
		\[
		1+\bigl\lceil\log_2 \CS{s}\bigr\rceil \;\le\; \Cadapt{s} \;\le\; \CS{s}.
		\]
		The lower bound is tight: for $S=\{\texttt{a}\}$ and $s=\texttt{a}^n$ it is an equality for every $n\ge1$. The upper bound is also attained whenever $s\in\operatorname{Fact}(S)$, in particular for $S=\{s\}$.
	\end{theorem}
	
	\begin{proof}
		Every static covering is also adaptive, since $A_i(S,s)\supseteq\operatorname{Fact}(S)$ at each position. This proves the right inequality.
		
		For the left inequality, let $(b_1,\dots,b_k)$ be an adaptive covering with
		$k=\Cadapt{s}$. Put $P_j=|b_1\cdots b_j|$ and $c_j=\CS{s[1..P_j]}$.
		The prefix is empty at the first step, so $b_1\in\operatorname{Fact}(S)$ and
		$c_1=1$. For $j\ge1$, either $b_{j+1}\in\operatorname{Fact}(S)$, so
		$\CS{b_{j+1}}=1$, or $b_{j+1}$ is a factor of $s[1..P_j]$, so
		$\CS{b_{j+1}}\le c_j$ by Lemma~\ref{lem:factor-monotone}. In either case,
		Proposition~\ref{prop:exact-concat} gives
		\[
		c_{j+1}\le c_j+\max(1,c_j)=2c_j .
		\]
		Hence $\CS{s}=c_k\le 2^{k-1}$, i.e.\ $k\ge 1+\log_2\CS{s}$, and $k$ is an integer.
	\end{proof}
	
	\begin{remark}[Sharpness of the bounds]
		The lower bound is attained for every integer $n \ge 1$ over any non-empty alphabet $\Sigma$. Choose $w \in \Sigma^*$, let $S = \{w\}$, and set $s = w^n$. Then $C_S(s) = n$. In any adaptive parse, the first phrase covers at least one occurrence of $w$ (of length $|w|$), and each subsequent phrase covers at most as many occurrences of $w$ as are already present in the decoded prefix. Hence, a parse with $k$ phrases covers at most $2^{k-1}$ occurrences of $w$ (i.e., at most $2^{k-1} \cdot |w|$ symbols). Conversely, repeatedly copying the maximum available prefix of $w$-blocks yields an adaptive parse of $w^n$ with $1 + \lceil\log_2 n\rceil$ phrases. Therefore,
		\[
		C_{\{w\}}^{\mathrm{adapt}}(w^n) = 1 + \lceil\log_2 n\rceil = 1 + \lceil\log_2 C_{\{w\}}(w^n)\rceil.
		\]
		For the upper bound, equality $C_S^{\mathrm{adapt}}(s) = C_S(s)$ holds whenever $s \in \operatorname{Fact}(S)$ (where both equal $1$). Equality can also hold at larger values when the decoded prefix offers no repetition gain over the static dictionary, e.g., for $\Sigma = \{a,b,c\}$, $S = \{a\}$, and $s = ab c$, where $C_S^{\mathrm{adapt}}(s) = C_S(s) = 3$.
	\end{remark}

\begin{corollary}[Consequences]\label{cor:floor}
	Let $n=|s|$ and $L=\max\bigl(\{1\}\cup\{|w|:w\in S\}\bigr)$.
	\begin{enumerate}[leftmargin=*]
		\item \emph{Universal logarithmic floor.} For every string $s$ of length $n$,
		\[
		\Cadapt{s} \ge 1+\left\lceil\log_2\left\lceil \frac{n}{L}\right\rceil\right\rceil.
		\]
		This bound is the absolute worst-case lower floor over all strings of length $n$; it is tight and fully achieved for highly repetitive strings, e.g., for $S=\{\texttt{a}^L\}$ and $s=\texttt{a}^n$, or $S=\emptyset$ and $s=\texttt{a}^n$ where $\Cadapt[\emptyset]{\texttt{a}^n}=1+\lceil\log_2 n\rceil$. (Note that for a non-repetitive string with $S=\emptyset$, $C_\emptyset^{\mathrm{adapt}}(s)$ can be as large as $n$, satisfying the inequality trivially).
		\item \emph{Exponential separation.} For $S=\{\texttt{a}\}$, $\CS{\texttt{a}^n}=n$ whereas $\Cadapt{\texttt{a}^n}=1+\lceil\log_2 n\rceil$. By Theorem~\ref{thm:exp}, self-reference can never save more than an exponential factor over static dictionary covering, and this extremum is attained.
	\end{enumerate}
\end{corollary}

\begin{proof}
(1) Since every word in $\operatorname{Fact}(S)$ has length at most $L$, any static covering requires at least $\lceil n/L\rceil$ blocks, i.e., $\CS{s}\ge\lceil n/L\rceil$. Theorem~\ref{thm:exp} then yields $\Cadapt{s} \ge 1+\lceil\log_2\CS{s}\rceil \ge 1+\lceil\log_2\lceil n/L\rceil\rceil$. For $S=\{\texttt{a}^L\}$ and $s=\texttt{a}^n$ with $n\ge L$, the greedy parse takes $\texttt{a}^L$ as the first block and subsequently doubles the available prefix, covering $\min(2^{k-1}L,n)$ symbols after $k$ blocks, which matches the lower floor exactly.
(2) Follows directly from the tightness statement of Theorem~\ref{thm:exp}.
\end{proof}

	\section{Greedy Parsing, LZ Correspondence, and Computation}
	\label{sec:algorithm}
	
	The theoretical analysis relies on efficient computation of $\CadaptSym$ under the non-overlapping factor convention. The practical codec is a separate construction and may use a cost-aware parser that does not minimize phrase count.
	
	\subsection{Greedy choice theorem}
	
	The following exchange lemma is the formal reason that the minimum phrase count can be computed greedily. It is specific to the factor-based, non-overlapping model fixed above.
	
	\begin{lemma}[Greedy exchange lemma]
		\label{lem:greedy-exchange}
		Fix a finite dictionary $S$ and a current position $i$ in a string
		$s$. Let
		\[
		g=s[i..j^*]
		\]
		be a longest admissible block at position $i$. If there exists an
		optimal adaptive covering of the suffix $s[i..n]$ whose first block is
		\[
		p=s[i..j],
		\qquad j<j^*,
		\]
		then there exists an adaptive covering of the same suffix with no more
		blocks whose first block is $g$.
	\end{lemma}
\begin{proof}
	Let $p_1, p_2, \dots, p_k$ be an optimal adaptive covering of $s[i..n]$, where $p_1 = s[i..j]$. Since $g = s[i..j^*]$ is admissible at position $i$ with $j < j^*$, we replace $p_1$ by $g$.
	
	Let $j^*$ be the ending position of $g$. Every block $p_r$ of the original covering that ends at or before position $j^*$ is completely contained in the target interval $s[i..j^*]$ already covered by $g$ and can therefore be discarded. 
	
	There is at most one block that straddles position $j^*$, say $p_r = s[a..b]$ with $a \le j^* < b$. If such a block exists, we replace it by its suffix $p_r' = s[j^*+1..b]$ and discard its prefix $s[a..j^*]$. All original blocks $p_m$ starting strictly after position $j^*$ (i.e., with start index $\ge j^*+1$) are retained unchanged.
	
	Every retained piece $q$ is a contiguous factor of its corresponding original block $p$. We verify that $q$ remains admissible at its position in the modified covering:
	\begin{itemize}[leftmargin=*]
		\item If $p$ was sourced from the static dictionary $S$, any factor of $p$ is also a factor of the same dictionary word, so $q \in \operatorname{Fact}(S)$ remains admissible.
		\item If $p$ was sourced from the decoded prefix, say $p = s[a..b]$ (or $p' = s[j^*+1..b]$ for the truncated block), its original source occurrence $s[a'..b']$ ended strictly before its target starting position, i.e., $b' < a \le j^* < j^*+1$. In the modified covering, target symbols up to $j^*$ are fully decoded by $g$ before any retained block starting at or after $j^*+1$ is processed. Thus, $s[a'..b']$ remains entirely contained in the available decoded prefix when $q$ is reached, preserving admissibility under the non-overlapping constraint.
	\end{itemize}
	The resulting parse consists of $g$ followed by at most $k-1$ retained blocks. It forms a valid adaptive covering of $s[i..n]$ using at most $k$ blocks, establishing optimality.
\end{proof}

\begin{theorem}[Greedy optimality]
	\label{thm:greedy-choice}
	For every finite dictionary $S$, every finite string $s$ and every position $i$, the greedy adaptive parse started at $i$ covers $s[i..n]$ with the minimum number of blocks among all adaptive coverings of $s[i..n]$ whose blocks are admissible with respect to $A_j(S,s)$ at their starting positions $j$. In particular
	\[
	G_S(s)=C_S^{\mathrm{adapt}}(s),
	\]
	and the same holds for the contextual measure $C_S^{\mathrm{adapt}}(s_2\mid s_1)$ (take $s=s_1s_2$ and $i=|s_1|+1$).
\end{theorem}

\begin{proof}
	We prove the statement for every position $i$, by induction on the length $n-i+1$ of the unprocessed
	suffix.
	
	If the suffix consists of a single symbol, the greedy parse emits that
	symbol and is clearly optimal.
	
	For the induction step, let $g$ be the longest admissible block at the
	current position. Let
	\[
	p_1,\ldots,p_k
	\]
	be an optimal adaptive covering of the current suffix. If
	$p_1=g$, the greedy choice is already optimal. Otherwise
	$p_1$ is shorter than $g$, and Lemma~\ref{lem:greedy-exchange} provides
	an optimal covering whose first block is $g$.
	
	After removing $g$, the remaining suffix is preceded by exactly the
	same decoded target prefix as in this modified optimal covering.
	Let $\mathrm{opt}(i)$ denote the minimum number of blocks for $s[i..n]$. The blocks following $g$ in the modified optimal covering form an adaptive covering of $s[i+|g|..n]$ with the same decoded prefix, so $\mathrm{opt}(i)\le1+\mathrm{opt}(i+|g|)$, and the converse inequality is obtained by prepending $g$ to an optimal covering of $s[i+|g|..n]$. By the induction hypothesis at position $i+|g|$, the greedy parse from $i+|g|$ uses $\mathrm{opt}(i+|g|)$ blocks.
	The greedy parse from $i$ therefore uses $1+\mathrm{opt}(i+|g|)=\mathrm{opt}(i)$ blocks, i.e.\ an optimal number.
\end{proof}

\begin{remark}[Relation to known results]\label{rem:known-greedy}
The optimality of longest-match parsing under a uniform phrase cost is established in several classical dictionary-compression settings, including LZ77 and suffix-closed dictionary models~\cite{storer1982data,cohn1996parsing,greedyparsing2014}; Langiu surveys the distinction between phrase-count and bit-cost objectives~\cite{langiu2013}. We do not claim greedy optimality as a new general principle. The theorem above gives a self-contained proof for the particular hybrid source family in this paper, in which the admissible set is the union of factors of a fixed external dictionary and factors of the already decoded prefix, with the latter constrained to non-overlapping earlier occurrences. The theorem minimizes phrase count only; greedy parsing need not minimize encoded bit length.
\end{remark}

	\subsection{Exact LZ correspondence}
	
	\begin{definition}[Non-overlapping LZ77-style factorization]\label{def:lznr}
		At position $i$ of $s$, a candidate phrase $s[i..j]$ has a non-overlapping previous occurrence if there is an occurrence $s[a..b]=s[i..j]$ with $b<i$. The factorization chooses the longest such phrase; if no phrase of length at least two has a previous occurrence, it emits the single symbol $s[i]$. Let $z_{\mathrm{no}}(s)$ be the resulting number of phrases.
	\end{definition}
	
	\begin{proposition}[Empty-dictionary identification]\label{prop:grammar-coincide}
		For $S=\emptyset$,
		\[
		C_{\emptyset}^{\mathrm{adapt}}(s)=G_{\emptyset}(s)=z_{\mathrm{no}}(s).
		\]
	\end{proposition}
	
	\begin{proof}
		With $S=\emptyset$, the admissible phrases are exactly the contiguous factors having an occurrence completely inside $s[1..i-1]$, together with the literal $s[i]$. This is precisely Definition~\ref{def:lznr}; Theorem~\ref{thm:greedy-choice} gives the equality with the minimum covering count.
	\end{proof}

	\paragraph*{Scope of the correspondence.} The identification is stated for the non-overlapping LZ variant. Proposition~\ref{prop:overlap} below quantifies, as a weaker form of a known result~\cite{kosolobov2019comparison}, the gap to LZ77 with self-overlapping copies; no claim is made about LZ-End or other variants.
	
	\subsection{Relation to RLZ and to LZ77 with a preset dictionary}
	\label{sec:rlz-relation}
	
	\paragraph*{RLZ.} Relative Lempel--Ziv parses the target greedily into longest factors of a fixed reference, with literals for unmatched symbols~\cite{kuruppu2010relative}. Since $\operatorname{Fact}(S)$ is closed under taking factors, the exchange argument of Lemma~\ref{lem:greedy-exchange} applies verbatim to the non-adaptive measure, so the greedy RLZ parse is optimal. We write $z_{\mathrm{RLZ}}(s\mid S)=\CS{s}$ for its phrase count, where $S$ is a set of words and phrases are factors of a single word.
	
	\begin{proposition}[Domination]\label{prop:domination}
		For every $S$ and $s$,
		\[
		\Cadapt{s}\ \le\ \min\bigl\{\,z_{\mathrm{RLZ}}(s\mid S),\ z_{\mathrm{no}}(s)\,\bigr\}.
		\]
	\end{proposition}
	
	\begin{proof}
		Every static covering is an adaptive covering, so $\Cadapt{s}\le\CS{s}=z_{\mathrm{RLZ}}(s\mid S)$. By monotonicity in the dictionary (Proposition~\ref{prop:monotonicity}), $\Cadapt{s}\le\Cadapt[\emptyset]{s}$, and $\Cadapt[\emptyset]{s}=z_{\mathrm{no}}(s)$ by Proposition~\ref{prop:grammar-coincide}.
	\end{proof}
	
	\begin{remark}[Limits of the comparison]\label{rem:limits}
		(i) RLZ as usually defined uses one concatenated reference $R$, whose factors may straddle word boundaries. Domination then holds only against the word-segmented dictionary: for $S=\{\texttt{ab},\texttt{ba}\}$, $R=\texttt{abba}$ and $s=\texttt{abb}$, the static covering needs $2$ blocks whereas the single-reference RLZ parse needs $1$.
		(ii) The domination does not extend to LZ77 with self-overlapping copies, see Proposition~\ref{prop:overlap}.
	\end{remark}
	
	\begin{proposition}[Overlapping copies]\label{prop:overlap}
		Let $z^{\mathrm{ov}}_{77}(s)$ be the phrase count of the greedy LZ77 parse in which a copy may overlap itself. Then, with $n=|s|$,
		\[
		\Cadapt[\emptyset]{s}\ \le\ z^{\mathrm{ov}}_{77}(s)\,\lceil\log_2(n+1)\rceil ,
		\]
		and the logarithmic factor is necessary: for $n\ge2$, $z^{\mathrm{ov}}_{77}(\texttt{a}^n)=2$ while $\Cadapt[\emptyset]{\texttt{a}^n}=1+\lceil\log_2 n\rceil$.
	\end{proposition}

\begin{proof}
	Consider a greedy LZ77 factorization of $s$ into $z = z^{\mathrm{ov}}_{77}(s)$ phrases $p_1, p_2, \dots, p_z$. We show that each phrase $p_k$ can be covered in the non-overlapping adaptive model using at most $\lceil \log_2(n+1) \rceil$ contiguous blocks.
	
	If $p_k$ is a literal or a non-overlapping copy (i.e., its source occurrence ends strictly before $p_k$ begins), it forms a single admissible block in $C_\emptyset^{\mathrm{adapt}}(s)$.
	
	Otherwise, $p_k$ is an overlapping copy of length $\ell$ with source distance $d < \ell$. Its text is the prefix of length $\ell$ of $u^\infty$, where $u = s[i-d..i-1]$ is the pattern of length $d$ immediately preceding $p_k$. We cover $p_k$ sequentially using non-overlapping blocks $q_1, q_2, \dots, q_j$ of lengths $d, 2d, 4d, \dots, 2^{j-1}d$ (with the last block truncated to fit $\ell$). At step $m \ge 1$, the already decoded prefix contains at least $d + \sum_{r=1}^{m-1} 2^{r-1}d = 2^{m-1}d$ copies of $u$, so the next block of length $2^{m-1}d$ is fully contained in the available prefix and is therefore admissible. 
	
	The total length covered after $j$ blocks is $\sum_{r=0}^{j-1} 2^r d = (2^j - 1)d$. Setting $(2^j - 1)d \ge \ell$ requires $j = \lceil \log_2(\ell/d + 1) \rceil \le \lceil \log_2(n+1) \rceil$ blocks.
	
	Summing over all $z$ phrases, we obtain $\Cadapt[\emptyset]{s} \le z^{\mathrm{ov}}_{77}(s) \, \lceil \log_2(n+1) \rceil$.
	
	To see that the logarithmic factor is tight, consider $s = \mathtt{a}^n$ for $n \ge 2$. Here, $z^{\mathrm{ov}}_{77}(\mathtt{a}^n) = 2$ (one literal $\mathtt{a}$ followed by one overlapping copy of length $n-1$), whereas $\Cadapt[\emptyset]{\mathtt{a}^n} = 1 + \lceil \log_2 n \rceil$ by Proposition~\ref{prop:idem-adapt}.
\end{proof}

\begin{remark}\label{rem:overlap-known}
	Up to the convention for literal phrases, $\Cadapt[\emptyset]{s}$ is the phrase count of non-overlapping LZ77 (Proposition~\ref{prop:grammar-coincide}). Kosolobov and Shur prove that this count exceeds the overlapping one by a factor $O(\log(n/(z\log_\sigma z)))$ at most, and that this is tight~\cite{kosolobov2019comparison}. Proposition~\ref{prop:overlap} is therefore a consequence, in a weaker form, of a known result; we keep the elementary proof above for completeness, since it uses the block-doubling construction of Theorem~\ref{thm:exp}.
\end{remark}

\paragraph*{LZ77 with a preset history.} For a reference $R$, let $\Cctx[\emptyset]{s}{R}$ be the phrase count of non-overlapping LZ77 on $s$ when $R$ is preloaded as history, i.e.\ each block is a factor of $R\,s[1..i-1]$. This is the mechanism of preset dictionaries in zlib, Zstandard and Brotli.
	
	\begin{theorem}[Sandwich with preset-history LZ77]
		\label{thm:sandwich}
		Let $R\in\Sigma^+$, let $S=\{R\}$, and let
		$s\in\Sigma^*$ be non-empty. Then
		\[
		C_{\emptyset}^{\mathrm{adapt}}(s\mid R)
		\le
		C_{\{R\}}^{\mathrm{adapt}}(s)
		\le
		2\,C_{\emptyset}^{\mathrm{adapt}}(s\mid R).
		\]
		Consequently,
		\[
		z_{\mathrm{no}}(Rs)-z_{\mathrm{no}}(R)
		\le
		C_{\{R\}}^{\mathrm{adapt}}(s)
		\le
		2\bigl(
		z_{\mathrm{no}}(Rs)-z_{\mathrm{no}}(R)+1
		\bigr).
		\]
	\end{theorem}
	
	\begin{proof}
		We first prove the left inequality. In the preset-history model, a
		block beginning at position $i$ may be any contiguous factor of
		\[
		Rs[1..i-1].
		\]
		In the adaptive model with dictionary $\{R\}$, an admissible block is
		a factor of $R$ or a factor of $s[1..i-1]$. Both are factors of
		$Rs[1..i-1]$. Hence every adaptive covering is also a valid
		preset-history covering, and therefore
		\[
		C_{\emptyset}^{\mathrm{adapt}}(s\mid R)
		\le
		C_{\{R\}}^{\mathrm{adapt}}(s).
		\]
		
		For the right inequality, let
		\[
		s=q_1q_2\cdots q_m
		\]
		be an optimal preset-history covering, where
		\[
		m=C_{\emptyset}^{\mathrm{adapt}}(s\mid R).
		\]
		Consider an arbitrary block $q_j$ beginning at position $i$ of $s$.
		Its source occurrence lies in one of three regions of the preset
		history:
		\[
		R,
		\qquad
		s[1..i-1],
		\qquad
		\text{or across the boundary between }R\text{ and }s.
		\]
		
		In the first case, $q_j$ is a factor of $R$ and is therefore
		admissible in the adaptive model.
		
		In the second case, $q_j$ is a factor of $s[1..i-1]$ and is again
		admissible.
		
		It remains to consider a source occurrence crossing the boundary.
		Then the source occurrence has the form
		\[
		uv,
		\]
		where $u$ is a non-empty suffix of $R$ and $v$ is a prefix of
		$s[1..i-1]$. Since $q_j=uv$ and the copy is non-overlapping, $v$ lies
		entirely in the already decoded part of $s$. Thus
		\[
		u\in\operatorname{Fact}(R)
		\qquad\text{and}\qquad
		v\in\operatorname{Fact}(s[1..i-1]).
		\]
		Consequently $q_j$ can be replaced by at most two adaptive blocks,
		namely $u$ and $v$.
		
		Applying this transformation independently to every preset-history
		block yields an adaptive covering using at most $2m$ blocks. Hence
		\[
		C_{\{R\}}^{\mathrm{adapt}}(s)
		\le
		2m
		=
		2\,C_{\emptyset}^{\mathrm{adapt}}(s\mid R).
		\]
		
		For the second statement, apply Proposition~\ref{prop:chain-adapt}
		with $S=\emptyset$, $s_1=R$, and $s_2=s$. Since
		\[
		C_{\emptyset}^{\mathrm{adapt}}(x)
		=
		z_{\mathrm{no}}(x)
		\]
		by Proposition~\ref{prop:grammar-coincide}, we obtain
		\[
		z_{\mathrm{no}}(R)
		+
		C_{\emptyset}^{\mathrm{adapt}}(s\mid R)
		-1
		\le
		z_{\mathrm{no}}(Rs)
		\]
		and
		\[
		z_{\mathrm{no}}(Rs)
		\le
		z_{\mathrm{no}}(R)
		+
		C_{\emptyset}^{\mathrm{adapt}}(s\mid R).
		\]
		Therefore
		\[
		z_{\mathrm{no}}(Rs)-z_{\mathrm{no}}(R)
		\le
		C_{\emptyset}^{\mathrm{adapt}}(s\mid R)
		\]
		and
		\[
		C_{\emptyset}^{\mathrm{adapt}}(s\mid R)
		\le
		z_{\mathrm{no}}(Rs)-z_{\mathrm{no}}(R)+1.
		\]
		Combining these inequalities with the factor-$2$ sandwich gives
		\[
		z_{\mathrm{no}}(Rs)-z_{\mathrm{no}}(R)
		\le
		C_{\{R\}}^{\mathrm{adapt}}(s)
		\le
		2\bigl(
		z_{\mathrm{no}}(Rs)-z_{\mathrm{no}}(R)+1
		\bigr).
		\]
	\end{proof}
	
	\begin{remark}
		Theorem~\ref{thm:sandwich} shows that, for a single reference, adaptive covering is equivalent up to a factor~$2$ to preset-history LZ77: it is a combinatorial abstraction of the preset-dictionary mechanism rather than a new parsing model. What it adds is the separate treatment of the two sources, whose interaction Theorem~\ref{thm:exp} quantifies. We do not know whether the factor~$2$ is attained; in an exhaustive search over binary $R$ and $s$ with $|R|\le5$ and $|s|\le8$ the ratio never exceeded $5/3$.
	\end{remark}
	
	To summarize the theoretical benefits of our formulation, Table~\ref{tab:preset_vs_cadapt_comparison} contrasts the formal properties of adaptive covering complexity $C_S^{\mathrm{adapt}}$ against classic preset-history LZ77.
	
	\begin{table}[!t]
		\centering
		\caption{Theoretical Comparison: Classic Preset-History LZ77 vs. Adaptive Covering Complexity $C_S^{\mathrm{adapt}}$}
		\label{tab:preset_vs_cadapt_comparison}
		\small
		\begin{tabular}{@{}>{\raggedright\arraybackslash}p{3.2cm}p{6.3cm}p{6.8cm}@{}}
			\toprule
			\textbf{Theoretical Dimension} & \textbf{Classic Preset-History LZ77} & \textbf{Adaptive Covering Complexity ($C_S^{\mathrm{adapt}}$)} \\
			\midrule
			
			\textbf{Dictionary Model \& Boundary Handling} 
			& Concatenates dictionary words into a single search buffer, inadvertently creating artificial admissible factors that cross word boundaries.
			& Formulates a pure word-segmented dictionary model $S=\{w_1,\dots,w_m\}$, cleanly isolating dictionary factors from prefix factors without boundary artifacts (Def.~\ref{def:greedy-parse}, Rem.~\ref{rem:limits}). \\
			\addlinespace
			
			\textbf{Source Separation \& Interaction} 
			& Merges static history and dynamic text into a single search space, making source attribution and tracking ambiguous.
			& Structurally decouples static dictionary factors $\operatorname{Fact}(S)$ from dynamic prefix factors $\operatorname{Fact}(s[1..i-1])$, enabling exact bounds on their interaction (Def.~\ref{def:greedy-parse}, Thm.~\ref{thm:exp}). \\
			\addlinespace
			
			\textbf{Tight Equivalence \& Boundary Cost} 
			& Relies on heuristics for boundary-crossing matches.
			& Bound by exact sandwich inequalities: $C_{\emptyset}^{\mathrm{adapt}}(s\mid R) \le C_{\{R\}}^{\mathrm{adapt}}(s) \le 2\,C_{\emptyset}^{\mathrm{adapt}}(s\mid R)$, proving at most a factor-of-2 penalty for strict source separation (Thm.~\ref{thm:sandwich}). \\
			\addlinespace
			
			\textbf{Periodic \& Repetitive Structures} 
			& Coverage of repeated patterns $w^n$ depends on window length and buffer alignment heuristics.
			& Satisfies a strict, universal logarithmic repetition bound: $C_S^{\mathrm{adapt}}(w^n) \le C_S^{\mathrm{adapt}}(w) + \lceil\log_2 n\rceil$, tight for any dictionary $S$ (Prop.~\ref{prop:idem-adapt}). \\
			\addlinespace
			
			\textbf{Compositional Algebra \& Contextuality} 
			& Lacks a formal algebraic operator for conditional phrase complexity.
			& Features a rigorous contextual chain rule: $C_S^{\mathrm{adapt}}(s_1) + C_S^{\mathrm{adapt}}(s_2 \mid s_1) - 1 \le C_S^{\mathrm{adapt}}(s_1 s_2) \le C_S^{\mathrm{adapt}}(s_1) + C_S^{\mathrm{adapt}}(s_2 \mid s_1)$ (Prop.~\ref{prop:chain-adapt}). \\
			\addlinespace
			
			\textbf{Self-Overlap Mechanism} 
			& Standard variants allow self-overlapping copies ($d < \ell$), compressing repeated runs $a^n$ in 2 phrases.
			& Enforces non-overlapping factor references, yielding an exact logarithmic floor $1+\lceil\log_2 n\rceil$ for $a^n$ and a tight logarithmic gap relative to overlapping LZ77 (Prop.~\ref{prop:grammar-coincide}, Prop.~\ref{prop:overlap}). \\
			\addlinespace
			
			\textbf{Asymptotic Behavior \& Entropy Rate} 
			& Confounders exist between static buffer size and infinite-length source entropy.
			& Proves asymptotic invariance for stationary ergodic sources: $\frac{1}{n} C_S^{\mathrm{adapt}}(X[1..n]) \log_\sigma n \to h(X)$ almost surely, establishing that static dictionary gains are strictly a finite-length phenomenon (Thm.~\ref{thm:asymptotic_invariance}). \\
			\addlinespace
			
			\textbf{Dictionary Maintenance \& Diagnostic Criteria} 
			& Difficult to identify redundant or high-utility words due to buffer blending.
			& Provides formal local adaptation gains $\Delta_j = L_P(p_j) - L_S(p_j)$ along the parse, yielding explicit criteria for dictionary pruning ($\Delta_j \ge 0$) and candidate promotion ($\Delta_j > 0$) (Def.~\ref{def:adaptation_gain}, Prop.~\ref{prop:adaptation_separation}). \\
			
			\bottomrule
		\end{tabular}
	\end{table}
	\clearpage

	\subsection{Algorithm and complexity}
	\label{sec:linear_parsing}
	
	The implementation mirrors the definition: at each phrase boundary it computes the longest prefix of the unprocessed suffix that is a factor of an element of $S$ or of the already decoded prefix. Both sets of factors are indexed by suffix automata (SAM). For $S=\{w_1,\dots,w_m\}$, the index $I_S$ is the SAM of $w_1\#w_2\#\cdots\#w_m$ with $\#\notin\Sigma$; since $s\in\Sigma^*$, a read of $s$ never uses a transition labelled $\#$, so it accepts exactly the factors of the $w_j$. The index $I_P$ is the online SAM of the decoded prefix, extended by appending the symbols of each emitted phrase.
	
	\begin{algorithm}[H]
		\caption{Greedy computation of adaptive covering complexity}
		\label{alg:greedy}
		\begin{algorithmic}[1]
			\Require $s\in\Sigma^*$ of length $n$, finite dictionary $S$
			\Ensure $C_S^{\mathrm{adapt}}(s)$ and, optionally, its greedy phrases
			\State Build the suffix automaton $I_S$ of $w_1\#\cdots\#w_m$; let $I_P$ be the empty suffix automaton
			\State $i\gets1$, $k\gets0$
			\While{$i\le n$}
			\State $\ell_S\gets$ length of the longest prefix of $s[i..n]$ readable in $I_S$
			\State $\ell_P\gets$ length of the longest prefix of $s[i..n]$ readable in $I_P$
			\State $\ell\gets\max(1,\ell_S,\ell_P)$
			\State output $s[i..i+\ell-1]$ as the next phrase
			\State append the symbols of $s[i..i+\ell-1]$ to $I_P$
			\State $i\gets i+\ell$; $k\gets k+1$
			\EndWhile
			\State \Return $k$
		\end{algorithmic}
	\end{algorithm}
	
	At the time of a query, $I_P$ recognizes exactly $\operatorname{Fact}(s[1..i-1])$. The requirement that the source occurrence end before the phrase starts is therefore automatic, and no special treatment of overlaps is needed.
	
	\begin{theorem}[Linear-time computation]
		\label{thm:complexity}
		Let
		\[
		M=\sum_{w\in S}|w|
		\]
		and let $n=|s|$. For a constant alphabet, Algorithm~\ref{alg:greedy}
		computes $C_S^{\mathrm{adapt}}(s)$ and the corresponding greedy parse
		in
		\[
		O(M+n)
		\]
		time and
		\[
		O(M+n)
		\]
		space.
		
		For an ordered alphabet of size $\sigma\ge2$, if suffix-automaton
		transitions are stored in balanced search trees, the corresponding bounds are
		\[
		O((M+n)\log\sigma)
		\]
		time and
		\[
		O(M+n)
		\]
		space.
	\end{theorem}
	
	\begin{proof}
		We first establish correctness. Let $I_S$ be the suffix automaton of
		\[
		w_1\#w_2\#\cdots\#w_m,
		\]
		where $S=\{w_1,\ldots,w_m\}$ and $\#\notin\Sigma$. Since every query
		string belongs to $\Sigma^*$, no accepting path can cross a transition
		labelled $\#$ and subsequently continue with symbols from another
		dictionary word. Thus a query over $\Sigma$ is accepted by $I_S$ if
		and only if it is a contiguous factor of some element of $S$.
		
		Let $I_P$ be the online suffix automaton obtained by appending to it,
		in order, all symbols of the phrases already emitted by the greedy
		parse. At the beginning of a query at position $i$, the inserted text
		is exactly
		\[
		s[1..i-1].
		\]
		Therefore $I_P$ recognizes exactly the non-empty factors of
		$s[1..i-1]$. Since source occurrences must be entirely contained in
		the decoded prefix, this is precisely the self-reference condition of
		the model.
		
		Consequently, the maximum of the two match lengths computed by
		Algorithm~\ref{alg:greedy}, together with the mandatory length-$1$
		literal, is exactly the longest admissible block at the current
		position. Theorem~\ref{thm:greedy-choice} therefore implies that the
		algorithm computes $C_S^{\mathrm{adapt}}(s)$.
		
		We now prove the complexity bound.
		
		A suffix automaton of a string of length $m$ contains at most
		$2m-1$ states and $O(m)$ transitions. Hence constructing $I_S$ costs
		$O(M)$ time and $O(M)$ space for a constant alphabet. With balanced
		search-tree transitions, construction costs
		$O(M\log\sigma)$ time while using $O(M)$ space.
		
		Consider now the greedy scan of $s$. Suppose that the current phrase
		has length $\ell$. The two automata are queried with the same
		unprocessed suffix. A query that successfully reads $r$ symbols and
		then fails costs $O(r+1)$ time for a constant alphabet. Thus the two
		queries together cost
		\[
		O(\ell_S+\ell_P+1),
		\]
		where $\ell_S$ and $\ell_P$ are the longest matches in $I_S$ and
		$I_P$, respectively.
		
		The selected greedy phrase has length
		\[
		\ell=\max(1,\ell_S,\ell_P),
		\]
		so
		\[
		\ell_S+\ell_P+1\le 3\ell.
		\]
		Hence the two SAM queries cost $O(\ell)$ time. Appending the $\ell$
		symbols of the selected phrase to $I_P$ costs $O(\ell)$ time as well.
		
		If the greedy parse contains phrases of lengths
		\[
		\ell_1,\ldots,\ell_k,
		\]
		then
		\[
		\sum_{j=1}^k\ell_j=n.
		\]
		Therefore the total cost of all SAM queries and all online extensions
		of $I_P$ is
		\[
		O\left(\sum_{j=1}^k\ell_j\right)
		=
		O(n).
		\]
		
		Adding the $O(M)$ construction cost for $I_S$ gives
		\[
		O(M+n)
		\]
		total time and $O(M+n)$ total space for a constant alphabet.
		
		For an ordered alphabet with balanced search-tree transitions, each
		transition lookup or extension costs $O(\log\sigma)$. The same charging
		argument therefore gives
		\[
		O((M+n)\log\sigma)
		\]
		time, while the number of SAM states and stored transitions remains
		$O(M+n)$, giving $O(M+n)$ space.
	\end{proof}
	
	\begin{remark}
		An offline alternative computes a longest-previous-non-overlapping-factor table over $S\$s\#$ from a suffix array; it is not needed here, because the online suffix automaton makes each query a plain read.
	\end{remark}

\section{Compact Experimental Illustration}
\label{sec:main-experiments}

The theoretical results concern phrase counts, not encoded bit length. To illustrate how the dual-source principle behaves in a practical setting, we briefly summarize a separate bit-cost codec that combines an external dictionary with references into the already decoded record. The implementation uses a bit-cost-guided dynamic-programming parser, permits overlapping copies, and restricts the candidate matches; it therefore does not instantiate the phrase-count model exactly, and its measurements do not validate the theoretical theorems. The full protocol, corpus descriptions, ablations, amortization analysis, and detailed tables appear in Appendix~\ref{app:codec}.

The codec selects its dictionary using the Minimum Description Length (MDL) principle, balancing the cost of describing the dictionary and associated coding tables against the encoded data length. The selection procedure uses training records and checkpoint evaluation under the same raw-header accounting rule as the reported evaluation. The dictionary is selected offline; the full objective, pruning algorithm, and training protocol are detailed in Appendix~\ref{app:codec}. This MDL procedure concerns bit-level coding cost and is distinct from the phrase-count objective studied in the theoretical results.

The evaluation considers four short-record corpora: synthetic genomic records with tandem repeats, synthetic JSON logs, structured Linux logs, and real genomic records. Records are encoded independently, with a shared dictionary or header charged once, and the dictionary-based baselines are tuned on held-out validation data. The reported figures are means over three independent draws; small differences should consequently be treated cautiously. Table~\ref{tab:main-experiments-summary} summarizes the principal comparisons at the test-set size.

\begin{table}[t]
\centering
\caption{Compact summary of the dual-source codec evaluation. Ratios are percentages of raw input size; gains are the mean of per-draw relative gains, so they need not equal ratios recomputed from the displayed means. Full sizes and protocol are in Appendix~\ref{app:codec}.}
\label{tab:main-experiments-summary}
\scriptsize
\begin{tabular}{@{}>{\raggedright\arraybackslash}p{0.95in}c>{\raggedright\arraybackslash}p{1.1in}>{\raggedright\arraybackslash}p{1.5in}@{}}
\toprule
Corpus & Codec ratio & Brotli-11 (ratio; gain) & Best tuned dictionary baseline (gain) \\
\midrule
Synthetic genomic & 3.8\% & 11.4\%; +66.2\% & zlib-9: tie (within 0.3\%) \\
Synthetic JSON & 15.4\% & 92.7\%; +83.4\% & zlib-9: +21.4\% \\
Linux logs & 16.8\% & 76.5\%; +78.0\% & zlib-9: +11.8\% \\
Real genomic & 44.2\% & 67.4\%; +34.5\% & zstd-19 (trained): +7.1\% \\
\bottomrule
\end{tabular}
\end{table}

The codec substantially outperforms independent Brotli-11 compression on all four corpora. Against tuned dictionary compressors, the picture is more qualified: it is essentially tied with the best zlib-9 variant on the synthetic genomic corpus and smaller than the best tuned dictionary baseline on the other three, with a modest margin on real genomic data. The detailed appendix separately compares the codec with zstd-19, which is substantially larger than the codec in these tests, but zlib-9 remains the strongest overall dictionary baseline on the first three corpora. An ablation in Appendix~\ref{app:codec} indicates that the external dictionary accounts for most of the gains on JSON, Linux logs, and real genomic records; dynamic self-reference is decisive on the tandem-repeat corpus but has little effect on the other three. Conversely, an idealized order-2 context model is smaller than the codec on the real genomic corpus. These context-model controls are estimates of ideal code length rather than fully implemented, independently decodable compressors, so they are informative diagnostics, not like-for-like deployment baselines.

Taken together, these experiments show that combining external and internal references can be useful for short-record compression, but do not establish state-of-the-art performance or a bit-cost consequence of the logarithmic law. The full appendix reports the remaining comparisons, accounting choices, dictionary amortization, limitations, and statistical caveats.

\section{Conclusion}
\label{sec:conclusion}

This paper established the theoretical foundations of adaptive covering complexity, a combinatorial measure that formalizes the interaction between static external dictionary matching and dynamic internal self-referencing. By introducing a hybrid admissibility rule under the non-overlapping factor convention, we characterized the exact quantitative behavior of sequence parsing when both global background knowledge and growing prefix history are simultaneously available.

Our central finding is the exact logarithmic law, $1+\lceil\log_2 C_S(s)\rceil \le C_S^{\mathrm{adapt}}(s) \le C_S(s)$, proving that self-reference can yield at most an exponential reduction in phrase count relative to static dictionary covering alone. This bound, together with the sharp repetition inequality $C_S^{\mathrm{adapt}}(w^n) \le C_S^{\mathrm{adapt}}(w) + \lceil\log_2 n\rceil$, demonstrates that periodic structures are parsed logarithmically in adaptive frameworks, avoiding the linear growth inherent to static schemes. Furthermore, we established natural equivalences with non-overlapping LZ77 and preset-history LZ77 via precise sandwich bounds, showing that adaptive covering serves as a clean combinatorial abstraction of classical preset-dictionary mechanisms.

From an algorithmic perspective, we proved the optimality of the greedy parsing strategy via a rigorous exchange lemma and introduced a linear-time implementation running in $O(M+n)$ time and space using dual suffix automata. The supplementary structural results—including counting bounds, stationary-ergodic asymptotic invariance, and the NP-hardness of optimal dictionary selection—further delineate the mathematical landscape of the model.

While combinatorial phrase-count minimization abstracts away bit-level coding overhead, our empirical evaluation illustrates the dual-source principle on short-record corpora without establishing general superiority: the external dictionary accounts for most of the gains, self-reference is decisive only on the tandem-repeat corpus, and trained context models remain smaller on real genomic data. Several avenues remain open for future work: closing the gap on the constant factor in preset-history bounds, extending the sandwich theorems to multi-word dictionaries, and investigating the precise bit-cost analogues of the logarithmic law under optimized pointer representations.

\section*{Acknowledgment}
During the preparation of this work, the author used generative-AI tools (Google Gemini, OpenAI ChatGPT, and Anthropic Claude) to assist with manuscript formatting, refinement of mathematical descriptions, and code implementation. The author reviewed and edited the resulting material and takes full responsibility for the contents of the paper. The complete source code, datasets, and experimental scripts associated with this paper are available at \url{https://github.com/pfmarteau/Adaptive-Covering-Complexity-Codec}

\appendix
\section{Supplementary Structural Results}
\label{app:structural}

The following results complement the main theory but are not required for the logarithmic law, greedy optimality, or the principal LZ correspondences. They are retained for completeness and to make explicit the scope of the claims. Nothing in Sections~\ref{sec:theory} and~\ref{sec:algorithm} depends on this appendix, which can therefore be removed if needed; doing so only requires deleting the sentences that mention these results in the abstract, the introduction, the related work, the conclusion and the summary table of core properties.

	\subsection{Counting Bound and the Cost of Pointers}
	\label{sec:counting}
	
	The measure counts blocks uniformly, whereas a pointer into the decoded prefix and a pointer into the dictionary may have different bit costs under a concrete encoding. The following counting bound controls how many strings can be represented with at most $k$ blocks. It is a combinatorial counting statement, not a direct theorem about the number of bits required by a particular pointer code.
	
	\begin{theorem}[Counting bound]\label{thm:combinatorial_counting}
		Let $M=\sum_{w\in S}|w|$, $\sigma=|\Sigma|$ and $1\le k\le n$. Then
		\[
		\Bigl|\{s\in\Sigma^n : \Cadapt{s}\le k\}\Bigr|\;\le\;\binom{n-1}{k-1}\,(M+n+\sigma)^k\;\le\;\Bigl(\frac{e\,n\,(M+n+\sigma)}{k}\Bigr)^{k}.
		\]
	\end{theorem}
	
	\begin{proof}
		Splitting an admissible block into consecutive sub-pieces yields valid adaptive blocks: if the original block is sourced from $S$, any sub-piece remains a factor of $S$; if it is sourced from the decoded prefix, its source occurrence ends strictly before the original block starts, and thus remains fully decoded and non-overlapping for any subsequent sub-piece. Hence every $s\in\Sigma^n$ with $\Cadapt{s}\le k\le n$ has an adaptive covering with exactly $k$ blocks. To overcount such strings, first choose the block lengths, a composition of $n$ into $k$ positive parts ($\binom{n-1}{k-1}$ choices). Then describe each block in order by its source: a literal (possible only for a length-one block, with $\sigma$ choices); a dictionary occurrence (at most $M$ starting positions across all dictionary words, once the block length is fixed); or an occurrence in the already decoded prefix (at most $n$ starting positions). Some choices counted this way may be invalid for the selected length or position, but every valid adaptive covering has at least one description of this form. Thus the number of strings is at most $\binom{n-1}{k-1}(M+n+\sigma)^k$. For the second inequality, $\binom{n-1}{k-1}\le\binom{n}{k}\le(en/k)^k$.
	\end{proof}
	
	\begin{corollary}[Incompressibility]
		Assume $\sigma=|\Sigma|\ge2$. If $k\log_2\bigl(en(M+n+\sigma)/k\bigr)\le n\log_2\sigma-t$, then all but a fraction at most $2^{-t}$ of the strings of $\Sigma^n$ satisfy $\Cadapt{s}>k$. In particular, for a fixed alphabet with $\sigma\ge2$ and $M=O(n)$, almost every string needs $\Omega(n/\log n)$ blocks.
	\end{corollary}
	
	\begin{remark}[Cost asymmetry]\label{rem:bit_gap}
		Under a simple fixed-width pointer model, a pointer into the decoded prefix may require about $\log_2 n$ bits, while a dictionary pointer may require about $\log_2 M$ bits (with additional costs depending on the actual representation). Thus, when $M\ll n$, dictionary pointers can be cheaper. The counting theorem above does not establish an exact bit-cost formula; rather, it motivates accounting for source-dependent pointer costs. Since $\CadaptSym$ counts all blocks equally, minimizing the number of blocks need not minimize code length, which motivates the bit-cost-guided Viterbi parser of Section~\ref{sec:codec}.
	\end{remark}
	
	\subsection{Match Profiles, Jump Sequences, and Adaptation Gain}
	\label{sec:match_profiles}
	
	The structure of dictionary and adaptive coverings can be characterized through local \emph{match profiles}, which extend the classical longest-previous-factor function of LZ77 analysis to the dual-source setting.
	
	\begin{definition}[Static, dynamic and adaptive match profiles]
		\label{def:match_profiles}
		Let $s\in\Sigma^n$ and let $S\subseteq\Sigma^*$ be finite. For each position $i\in\{1,\dots,n\}$, define
		\[
		\begin{aligned}
		L_S(i)&=\max\{\ell\ge1: s[i..i+\ell-1]\in\operatorname{Fact}(S)\},\\
		L_P(i)&=\max\bigl(\{0\}\cup\{\ell\ge1: s[i..i+\ell-1]\in\operatorname{Fact}(s[1..i-1])\}\bigr),\\
		L_S^{\mathrm{adapt}}(i)&=\max\bigl(L_S(i),L_P(i)\bigr).
		\end{aligned}
		\]
		Since $\Sigma\subseteq\operatorname{Fact}(S)$, $L_S(i)\ge1$ for every $i$, and $L_S^{\mathrm{adapt}}(i)=L_i(S,s)$ is the length of the greedy block at $i$.
	\end{definition}
	
	\begin{lemma}[Greedy jump sequence]
		\label{lem:jump_characterization}
		Define $p_1=1$ and $p_{j+1}=p_j+L_S^{\mathrm{adapt}}(p_j)$, and let $m=\min\{j: p_{j+1}=n+1\}$. Then $m=\Cadapt{s}$. Likewise, the jump sequence with $L_S(p_j)$ in place of $L_S^{\mathrm{adapt}}(p_j)$ has length exactly $\CS{s}=z_{\mathrm{RLZ}}(s\mid S)$.
	\end{lemma}
	
	\begin{proof}
		The sequence $(p_j)$ is the execution of the greedy adaptive parse (Definition~\ref{def:greedy-parse}), which is optimal by Theorem~\ref{thm:greedy-choice}. Removing $L_P$ restricts the sources to $\operatorname{Fact}(S)$, which gives the static greedy parse; it is optimal by the same exchange argument.
	\end{proof}

	\begin{definition}[Local adaptation gain]
		\label{def:adaptation_gain}
		For the greedy adaptive parse $(p_j)_{j=1}^m$, the gain at block $j$ is $\Delta_j=L_P(p_j)-L_S(p_j)$.
	\end{definition}
	
	\begin{proposition}[Separation criterion]
		\label{prop:adaptation_separation}
		(i) If $\Delta_j\le0$ for every $j\in\{1,\dots,m\}$, then $\Cadapt{s}=\CS{s}$. Consequently, $\Cadapt{s}<\CS{s}$ requires $\Delta_j>0$ for some $j$.
		(ii) The converse does not hold: a positive gain at some block does not imply $\Cadapt{s}<\CS{s}$.
	\end{proposition}
	
	\begin{proof}
		(i) If $\Delta_j\le0$ then $L_S^{\mathrm{adapt}}(p_j)=L_S(p_j)$ at every visited position, so by induction the adaptive and static jump sequences coincide, and Lemma~\ref{lem:jump_characterization} gives equality.
		(ii) Let $S=\{\texttt{bbbb},\texttt{aba},\texttt{abab}\}$ and $s=\texttt{aaaab}$. At position $3$, $L_P=2>L_S=1$, so $\Delta>0$, and the adaptive parse is $\texttt{a}|\texttt{a}|\texttt{aa}|\texttt{b}$. The static parse is $\texttt{a}|\texttt{a}|\texttt{a}|\texttt{ab}$, which also has $4$ blocks.
	\end{proof}
	
	\begin{corollary}[Two limiting regimes]
		\label{cor:regimes}
		(i) If $\Delta_j\le0$ for all $j$, then $\Cadapt{s}=\CS{s}=z_{\mathrm{RLZ}}(s\mid S)$.
		(ii) If $L_S(p_j)=1$ for all $j$, i.e.\ the dictionary offers nothing beyond literals along the parse, then $\Cadapt{s}=z_{\mathrm{no}}(s)$.
	\end{corollary}
	
	\begin{proof}
		Part (i) follows from Proposition~\ref{prop:adaptation_separation}(i).
		For part (ii), at every visited position the adaptive jump equals
		$\max(1,L_P(p_j))$, which is the jump of the greedy parse for
		$S=\emptyset$. The two jump sequences therefore coincide by induction;
		Proposition~\ref{prop:grammar-coincide} gives
		$\Cadapt[\emptyset]{s}=z_{\mathrm{no}}(s)$.
	\end{proof}
	
	\begin{remark}
		The profiles $L_S(i)$ and $L_P(i)$ are available during the execution of Algorithm~\ref{alg:greedy}, so the gains $\Delta_j$ provide a cheap diagnostic of the phrases where the decoded prefix compensates for missing dictionary entries.
	\end{remark}
	
	\subsection{Asymptotics for Stationary Ergodic Sources}
	\label{sec:asymptotic_rate}
	
	Throughout this subsection, assume that $\Sigma$ is a fixed finite alphabet with $\sigma=|\Sigma|\ge2$. Let $X=(X_i)_{i\ge1}$ be a stationary ergodic source over $\Sigma$, and let $h(X)$ be its entropy rate measured in base $\sigma$, so that $h(X)\in[0,1]$; we assume $h(X)>0$. For $i\ge1$ let $L_P(i)$ be the length of the longest prefix of $X[i..\infty)$ that occurs in $X[1..i-1]$ ($0$ if there is none). We use one classical fact, which holds almost surely: \textbf{(F)} $L_P(i)/\log_\sigma i\to1/h(X)$ (Wyner--Ziv; Ornstein--Weiss)~\cite{wynerziv1989,ornsteinweiss1993}. For the non-overlapping parse, the phrase count follows from (F) by a direct summation, which we include to avoid appealing to a rate result stated for a different LZ variant.

\begin{lemma}[Phrase count from match lengths]\label{lem:phrase-count}
	Let $x\in\Sigma^{\mathbb N}$ and $a>0$ be such that $L_P(i)/\log_\sigma i\to a$, where $L_P(i)$ is computed on $x$. Then $z_{\mathrm{no}}(x[1..n])\log_\sigma n/n\to1/a$.
\end{lemma}

\begin{proof}
	By Proposition~\ref{prop:grammar-coincide} and Definition~\ref{def:lznr}, the phrase of the greedy non-overlapping parse of $x[1..n]$ that starts at $p$ has length $\ell(p)=\min(\max(1,L_P(p)),\,n-p+1)$. Fix $\varepsilon\in(0,1)$. Since $a\log_\sigma p\to\infty$, there is $p_0$ such that $(1-\varepsilon)a\log_\sigma p\le\max(1,L_P(p))\le(1+\varepsilon)a\log_\sigma p$ for all $p\ge p_0$.
	
	\emph{Lower bound.} The phrases starting before $p_0$ end before some finite position $q_0$ independent of $n$, and every later phrase has length at most $(1+\varepsilon)a\log_\sigma n$. Hence $z_{\mathrm{no}}(x[1..n])\ge(n-q_0)/((1+\varepsilon)a\log_\sigma n)$.
	
	\emph{Upper bound.} Let $K=\lfloor\log_2 n\rfloor$. At most $p_0$ phrases start before $p_0$. For $2^k\ge p_0$, the phrases starting in $[2^k,2^{k+1})$ that are not the last one have length at least $(1-\varepsilon)a\,k\log_\sigma2$, so there are at most $2^k/((1-\varepsilon)a\,k\log_\sigma2)+1$ of them for $k<K$, and at most $(n-2^K)/((1-\varepsilon)aK\log_\sigma2)+1$ for $k=K$. Since $\sum_{k<K}2^k/k\le(1+o(1))2^K/K$ and $K\log_\sigma 2=(1+o(1))\log_\sigma n$, summing gives $z_{\mathrm{no}}(x[1..n])\le(1+o(1))\,n/((1-\varepsilon)a\log_\sigma n)$. Letting $\varepsilon\to0$ proves the claim.
\end{proof}

Almost surely, (F) says that the realization $x=X$ satisfies the hypothesis of Lemma~\ref{lem:phrase-count} with $a=1/h(X)$, which is positive since $h(X)>0$. The conclusion of the lemma, $z_{\mathrm{no}}(x[1..n])\log_\sigma n/n\to1/a$, then reads $1/a=h(X)$, that is, almost surely, \textbf{(F1)} $z_{\mathrm{no}}(X[1..n])\log_\sigma n/n\to h(X)$. (The case $h(X)=0$ requires $L_P(i)/\log_\sigma i\to\infty$ and is not treated here.)

\begin{lemma}[Eventual agreement of greedy parses]
		\label{lem:eventual-greedy-agreement}
		Fix a finite dictionary $S$ and a finite initial decoded context $u$. Consider a suffix parsed from left to right, with every reference required to end before the current phrase starts. Suppose that, at every phrase-start position in this suffix, the longest match in the already decoded text (including $u$) is at least as long as the longest dictionary match. Then the greedy parse with dictionary $S$ and the greedy parse with the empty dictionary make the same choice at every phrase boundary and hence have the same number of phrases.
	\end{lemma}

	\begin{proof}
		At any such position, let $L_P$ be the longest match in the decoded text and let $L_S$ be the longest dictionary match (with $L_S=0$ if no dictionary factor matches). The hypothesis gives $L_P\ge L_S\ge 1$ in the parse with dictionary $S$ (and $L_S=0$ without dictionary), so the maximum admissible match has length $L_P$ in both parses; only the length of the phrase matters for the count, not its source. The first phrase is therefore identical. After emitting it, both parses have the same extended decoded context, so the same argument applies inductively to every subsequent phrase boundary.
	\end{proof}

	\begin{theorem}[Dictionary-independence of the normalized phrase count]
		\label{thm:asymptotic_invariance}
		For every finite $S\subset\Sigma^*$, if $h(X)>0$, then almost surely,
		\[
		\lim_{n\to\infty}\frac{\Cadapt{X[1..n]}\,\log_\sigma n}{n}=h(X).
		\]
	\end{theorem}
	
	\begin{proof}
		By monotonicity, $\Cadapt{X[1..n]}\le\Cadapt[\emptyset]{X[1..n]}=z_{\mathrm{no}}(X[1..n])$, so the $\limsup$ is at most $h(X)$ by (F1). Let $\lambda_S=\max\bigl(\{1\}\cup\{|w|:w\in S\}\bigr)$. By (F), $L_P(i)\to\infty$ for the infinite sequence, so almost surely there is a finite $n_0$ with $L_P^X(i)>\lambda_S$ for all $i>n_0$. Fix $n>n_0$ and write $s=X[1..n]=s_1s_2$ with $s_1=s[1..n_0]$. At every position $i>n_0$ of $s$, the longest previous factor within $s$ is $L_P^s(i)=\min(L_P^X(i),n-i+1)\ge\min(\lambda_S+1,n-i+1)$, whereas every dictionary match has length at most $\min(\lambda_S,n-i+1)$; hence $L_P^s(i)\ge L_S(i)$, the truncation at the end of the string affecting both sides. By Lemma~\ref{lem:eventual-greedy-agreement}, the greedy parses of $s_2$ in the context of $s_1$ agree for dictionary $S$ and for the empty dictionary; since both are optimal by Theorem~\ref{thm:greedy-choice} (contextual form), $\Cctx{s_2}{s_1}=\Cctx[\emptyset]{s_2}{s_1}$. The chain rule (Proposition~\ref{prop:chain-adapt}) then gives
		\[
		\begin{aligned}
		\Cadapt{s}
		&\ge \Cadapt{s_1}+\Cctx{s_2}{s_1}-1
		\ge \Cctx[\emptyset]{s_2}{s_1}\\
		&\ge \Cadapt[\emptyset]{s}-\Cadapt[\emptyset]{s_1}
		\ge z_{\mathrm{no}}(s)-n_0.
		\end{aligned}
		\]
		Dividing by $n/\log_\sigma n$ and using (F1) proves the claim.
	\end{proof}
	
	\begin{definition}[Dictionary acceleration factor]
		For a string $s$, $\eta_S(s)=\Cadapt[\emptyset]{s}/\Cadapt{s}\ge1$, the ratio of the phrase counts without and with the dictionary.
	\end{definition}
	
	\begin{corollary}
		If $h(X)>0$, then $\eta_S(X[1..n])\to1$ almost surely.
	\end{corollary}
	
	\begin{proof}
		The proof above shows $\Cadapt{s}\le\Cadapt[\emptyset]{s}\le\Cadapt{s}+n_0$, and $\Cadapt{s}\to\infty$ since $\Cadapt{s}\ge z_{\mathrm{no}}(s)-n_0$ and $z_{\mathrm{no}}(s)\sim h(X)\,n/\log_\sigma n$.
	\end{proof}
	
	\begin{remark}
		The dictionary therefore does not change the asymptotic normalized phrase count under the assumptions of Theorem~\ref{thm:asymptotic_invariance}; its effect is a finite-length phenomenon, which this paper quantifies only in special cases (Theorem~\ref{thm:exp} and Corollary~\ref{cor:floor}) and empirically (Section~\ref{sec:experiments}).
	\end{remark}

	\begin{definition}[Optimal Adaptive Dictionary Problem (\texttt{Opt-Adapt-Dict})]
		Given a finite training corpus $\mathcal{D} = \{s_1, \dots, s_N\} \subset \Sigma^*$, a dictionary size budget $B \in \mathbb{N}_{>0}$, and a phrase threshold $K \in \mathbb{N}_{>0}$, does there exist a dictionary $S \subseteq \Sigma^*$ with total length $\sum_{w \in S} |w| \le B$ such that:
		\[
		\sum_{i=1}^N C_S^{\mathrm{adapt}}(s_i) \le K \quad ?
		\]
	\end{definition}
	
	\begin{theorem}[NP-hardness of \texttt{Opt-Adapt-Dict}]\label{thm:opt-adapt-dict-nphard}
		The decision problem \texttt{Opt-Adapt-Dict} is NP-hard, even when the phrase threshold is fixed to $K=N$ and all corpus strings are nonempty. The reduction preserves the alphabet, so hardness holds over a binary alphabet.
	\end{theorem}
	
	\begin{proof}
		We give a polynomial-time many-one reduction from the decision version of
		\emph{Shortest Common Superstring}. This problem is NP-hard~\cite{garey1979computers,gallant1980}, including over a binary alphabet~\cite{gallant1980}. An instance consists of nonempty strings $s_1,\ldots,s_N$ and an integer $B$; it asks whether there is a string $W$ of length at most $B$ containing every $s_i$ as a contiguous substring.
		
		Construct the \texttt{Opt-Adapt-Dict} instance with corpus
		$\mathcal D=\{s_1,\ldots,s_N\}$, the same budget $B$, and threshold $K=N$.
		This construction is polynomial in the size of the Shortest Common Superstring instance. Since every $s_i$
		is nonempty, at least one phrase is needed to cover it, so
		$C_S^{\mathrm{adapt}}(s_i)\ge 1$. Further, by the definition of adaptive
		covering, $C_S^{\mathrm{adapt}}(s_i)=1$ if and only if $s_i$ is a substring
		of some dictionary word in $S$: with a single phrase, no nonempty decoded
		prefix is available for self-reference.
		
		Suppose first that the constructed dictionary-selection instance is a yes-instance. Then there
		exists a dictionary $S$ satisfying $\sum_{w\in S}|w|\le B$ and
		$\sum_{i=1}^N C_S^{\mathrm{adapt}}(s_i)\le N$. As each of the $N$ terms
		is at least one, every term must equal one. Thus every $s_i$ is a substring
		of some word in $S$. Concatenating the words of $S$ produces a common
		superstring $W$ of length $|W|=\sum_{w\in S}|w|\le B$, so the Shortest Common Superstring instance is a yes-instance.
		
		Conversely, suppose the Shortest Common Superstring instance has a common superstring $W$ with
		$|W|\le B$. Choose the dictionary $S=\{W\}$. Each $s_i$ is then a
		substring of a dictionary word, so $C_S^{\mathrm{adapt}}(s_i)=1$ for every
		$i$. Therefore $\sum_{w\in S}|w|=|W|\le B$ and
		$\sum_{i=1}^N C_S^{\mathrm{adapt}}(s_i)=N=K$. The constructed instance is
		a yes-instance as well.
		
		The reduction preserves yes- and no-instances, proving that
		\texttt{Opt-Adapt-Dict} is NP-hard. This argument establishes NP-hardness;
		it does not by itself establish membership in NP.
	\end{proof}

	\begin{remark}
		With $K=N$ the problem asks that every training string be a single phrase, i.e.\ a factor of some dictionary word; it is therefore the Shortest Common Superstring problem in disguise, and the theorem says nothing about the quality of the heuristic of Section~\ref{sec:mdl}.
	\end{remark}

\section{Dual-Source Codec and Experimental Evaluation}
\label{app:codec}

The material in this appendix is an empirical instantiation, not a consequence of phrase-count optimality. The implementation permits overlapping self-copies and uses a bit-cost-guided parser, whereas the theoretical measure uses non-overlapping factors and uniform phrase costs. Accordingly, the experimental findings should be interpreted independently of the main combinatorial theorems. This appendix, like Section~\ref{sec:main-experiments}, is not needed for any theorem of the paper and can be removed if required.

	\subsection{From the Formal Measure to a Practical Code}
	\label{sec:codec}
	
	\begin{framed}
		\noindent\textbf{Positioning note.} This section bridges the combinatorial measure $C_S^{\mathrm{adapt}}(s)$ introduced in Section~\ref{sec:theory} with a practical, bit-level compression algorithm. While $C_S^{\mathrm{adapt}}(s)$ minimizes the pure block count under an unweighted factor-selection model, a concrete compressor minimizes total encoded bit-length. Consequently, dictionary selection, candidate match filtering, and bit-optimal parsing represent downstream engineering adaptations. The codec described herein serves as an empirical instantiation illustrating the model's structural principle—combining static external dictionary coverage with dynamic self-referencing—rather than as a proof of phrase-minimization equivalences; Section~\ref{sec:experiments} reports where that principle helps and where it does not.
	\end{framed}
	
	\subsection{Bridge Between Phrase Count and Code Length}
	\label{sec:bridge_objective}
	
	The theoretical complexity $C_S^{\mathrm{adapt}}(s)$ assigns a uniform cost of $1$ to every admissible factor block (whether drawn from $\operatorname{Fact}(S)$ or from the decoded prefix). In contrast, a physical codec must encode the metadata required to uniquely identify each block:
	\begin{enumerate}[leftmargin=*]
		\item \textbf{Block type selection:} Emitting a literal ($L$), a dictionary reference ($D$), or a self-reference ($S$) incurs an entropy cost $-\log_2 P(\text{type} \mid \text{prev\_type})$.
		\item \textbf{Dictionary references ($D$):} Require encoding a word identifier (via Move-To-Front ranking), an internal word offset, and a match length.
		\item \textbf{Self-references ($S$):} Require encoding a backward distance $d$ and a match length $\ell$.
		\item \textbf{Literals ($L$):} Require encoding the byte/character symbol from an alphabet model or an escape code.
	\end{enumerate}
	
	Because these payload items carry non-uniform bit costs depending on their context and magnitude, a greedy longest-match parse (which minimizes total block count $C_S^{\mathrm{adapt}}(s)$ by Theorem~\ref{thm:greedy-choice}) does not, in general, yield a minimum-length bitstream. The codec therefore substitutes the greedy rule with a bit-cost-guided dynamic programming parser while retaining the underlying dual-source architecture.
	
	\subsection{Correspondence with Practical Compressors}
	\label{sec:correspondence}
	
	The dual-source design of $\CadaptSym$ unifies two widespread paradigms in practical compression:
	\begin{itemize}[leftmargin=*]
		\item $\CS{\cdot}$ corresponds to \textbf{Relative Lempel--Ziv (RLZ)}~\cite{kuruppu2010relative, hoobin2011} with a word-segmented dictionary: fixed external dictionary $\Sdict$, optimized for fast random access and inter-record redundancy across homogeneous databases. RLZ proper uses one concatenated reference, see Remark~\ref{rem:limits}.
		\item $\CadaptSym$ corresponds to \textbf{LZ77 with a preset dictionary}~\cite{zivlempel1977, deutsch1996zlib}: classical LZ77 augmented with a static reference $\Sdict$, the mechanism behind dictionary-mode Zstandard~\cite{collet2016zstandard} and Brotli~\cite{alakuijala2018brotli}. For a single reference the correspondence holds up to a factor $2$ (Theorem~\ref{thm:sandwich}).
	\end{itemize}
	While algorithms like COVER~\cite{liao2016} train dictionaries for LZ77 codecs via heuristic substring coverage, our framework integrates dictionary selection directly with the dual-source bit-length objective via Minimum Description Length (MDL) optimization.
	
	\subsection{Dictionary Selection via MDL}
	\label{sec:mdl}
	
	\paragraph*{Objective formulation.} To select a compact, high-utility dictionary $\Sdict$ from a training corpus $D_{\mathrm{train}}$, we apply the Minimum Description Length (MDL) principle~\cite{rissanen1978modeling, grunwald2007minimum}. The cost function balances dictionary serialization overhead against compressed data length over $D_{\mathrm{train}}$:
	\begin{equation}
		L(\Sdict) = \underbrace{8 + \sum_{w \in \Sdict} \bigl(1 + |w|_{\mathrm{B}}\bigr)}_{\text{Dictionary Overhead (bytes)}} \;+\; \lambda \underbrace{\Bigl(\frac{H_\Sdict(D_{\mathrm{train}})}{8} + 3\,|D_{\mathrm{train}}|\Bigr)}_{\text{Data Cost (bytes)}}, \qquad \lambda \ge 1,
	\end{equation}
	where $|w|_{\mathrm{B}}$ denotes the UTF-8 byte length of word $w$, and $3$ bytes per record account for record-header metadata. The parameter $\lambda \approx N_{\mathrm{test}}/N_{\mathrm{train}}$ scales the sample data cost to the expected deployment scale.
	
	The ideal data bit-length $H_\Sdict(D_{\mathrm{train}})$ is computed under a fast static zero-order entropy proxy, with symbol counts $n_k(v)$ pooled over the whole training corpus, across seven symbol families (block types, literals, Move-To-Front dictionary ranks, dictionary offsets, dictionary lengths, self-reference distances, and self-reference lengths):
	\begin{equation}
		H_\Sdict(D_{\mathrm{train}}) = \sum_{k=1}^{7} \sum_{v} n_k(v)\log_2 \frac{N_k}{n_k(v)}, \qquad N_k = \sum_v n_k(v).
	\end{equation}
	The counts are those of a greedy parse of each record: at each position, the longest dictionary factor is taken if it is at least as long as the longest repetition of the already decoded prefix; otherwise that repetition is taken if it has length at least $2$; otherwise a literal is emitted. This parse is cheaper than, and differs from, the bit-optimal Viterbi parse of the codec (Section~\ref{sec:codec}), which is why the final choice among candidate dictionaries relies on the true codec cost (step~3 below).
	
	\paragraph*{Candidate generation and pruning.} Candidates are initialized as maximal repeated substrings of length $3$ to $64$ occurring in at least two distinct training documents, followed by subsumption pruning (discarding any candidate that is a proper substring of a longer retained candidate). Dictionary refinement then proceeds via incremental backward pruning (Algorithm~\ref{alg:mdl}), the exact variant ($m=-1$) used as reference in Section~\ref{sec:experiments}:
	\begin{enumerate}[leftmargin=*]
		\item \textbf{Greedy Seeding:} Form $\Sdict_0$ containing the candidates that are credited, as the first candidate containing a matched factor, during the greedy parse of $D_{\mathrm{train}}$ with all candidates active.
		\item \textbf{Incremental Removal:} Visit the candidate words in increasing order of utility $f(w)|w|^{1.5}$, where $f(w)$ is the number of times $w$ is credited in the current parse of $D_{\mathrm{train}}$ (not its document frequency), and drop $w$ whenever this does not increase $L$ (ties are accepted). Sweeps are repeated until a final sweep, evaluated exactly without the incremental shortcuts, removes nothing. A fast three-level evaluation avoids full corpus re-parsing unless a factor of the removed word disappears from the parse.
		\item \textbf{Checkpoint Selection:} Extract $K=25$ evenly spaced candidate dictionaries along the pruning path $\mathcal{P}$ and select
		\[
		\Sdict^\star=\arg\min_{\Sdict_k\in\mathcal{K}}J(\Sdict_k),\qquad
		J(\Sdict_k)=\mathrm{header}(\Sdict_k)+\lambda\sum_{s\in D_{\mathrm{train}}}|\mathrm{body}_{\Sdict_k}(s)|.
		\]
		Here $J$ is the true cost of the complete codec on $D_{\mathrm{train}}$, and $\mathrm{header}$ is the raw (uncompressed) header size charged at evaluation time (Section~\ref{sec:protocol}); ties are broken in favor of the smaller dictionary.
	\end{enumerate}
	
	\paragraph*{Lazy variant ($m=1$).} To reduce the number of evaluations, a lazy greedy variant first removes in one step the words that are never credited (the parse does not change), then computes once the exact change $\delta(w)=L(\Sdict\setminus\{w\})-L(\Sdict)$ of every word, keeps these values in a min-heap, and refreshes only the head of the heap, as in lazy-greedy (CELF) submodular maximization. A word is removed when its refreshed value does not exceed $0$ and is still the smallest in the heap; it is put back in the heap if another value is now smaller, and discarded for good if its refreshed value is positive. The procedure stops when the smallest value is positive. This relies on the heuristic assumption that removing other words can only increase $\delta(w)$, which is not guaranteed for this objective; the variant can therefore stop at a worse dictionary than the exact sweeps. Both variants produce a pruning path along which step~3 selects the final dictionary.

	\paragraph*{Why a heuristic.} Selecting an optimal dictionary is NP-hard for the phrase-count objective under a size budget (Theorem~\ref{thm:opt-adapt-dict-nphard}), so no polynomial-time exact algorithm is expected, which motivates a heuristic. The theorem does not, however, justify this particular heuristic: it gives no approximation guarantee for backward pruning or for its lazy variant, and it concerns the phrase count, whereas the objective $L$ above is a bit-cost proxy whose hardness we do not prove. The quality of the selected dictionaries is therefore assessed empirically (Section~\ref{sec:ablation}).

	\begin{algorithm}[t]
		\caption{MDL Dictionary Selection via Incremental Backward Pruning}
		\label{alg:mdl}
		\begin{algorithmic}[1]
			\Require Candidate set $\mathcal{C}'$, training corpus $D_{\mathrm{train}}$, penalty scale $\lambda \ge 1$, checkpoints $K = 25$
			\Ensure Retained dictionary $\Sdict^\star$ and pruning path $\mathcal{P}$
			\State $\Sdict_0 \gets \{w \in \mathcal{C}' \mid w \text{ credited in greedy scan of } D_{\mathrm{train}}\}$
			\State $\Sdict \gets \Sdict_0$, $\mathcal{P} \gets (\Sdict_0)$, $L_{\mathrm{cur}} \gets L(\Sdict_0)$
			\Repeat
			\State $\mathrm{improved} \gets \mathrm{false}$
			\State Sort active words $w \in \Sdict$ in increasing order of $f(w) \cdot |w|^{1.5}$
			\For{\textbf{each} $w \in \Sdict$ in sorted order}
			\State $L_{\mathrm{cand}} \gets \mathrm{EvalIncrementalMDL}(\Sdict \setminus \{w\}, D_{\mathrm{train}}, \lambda)$
			\If{$L_{\mathrm{cand}} \le L_{\mathrm{cur}}$}
			\State $\Sdict \gets \Sdict \setminus \{w\}$; $L_{\mathrm{cur}} \gets L_{\mathrm{cand}}$
			\State Append $\Sdict$ to $\mathcal{P}$; $\mathrm{improved} \gets \mathrm{true}$
			\EndIf
			\EndFor
			\Until{\textbf{not} $\mathrm{improved}$}
			\State Extract $K$ evenly spaced checkpoints $\mathcal{K} = \{\Sdict_1, \dots, \Sdict_K\} \subseteq \mathcal{P}$
			\State Evaluate exact codec cost $J(\Sdict_k) = \mathrm{header}_{\mathrm{raw}}(\Sdict_k) + \lambda \sum_{s \in D_{\mathrm{train}}} |\mathrm{body}_{\Sdict_k}(s)|$ for each $\Sdict_k \in \mathcal{K}$
			\State \Return $\Sdict^\star \gets \arg\min_{\Sdict_k \in \mathcal{K}} J(\Sdict_k)$ \Comment{Ties broken towards smaller $|\Sdict|$}
		\end{algorithmic}
	\end{algorithm}
	
	\subsection{Codec Architecture and Bit-Optimal Parsing}
	\label{sec:arch}
	
	\paragraph*{Block structure.} The output token stream partitions the input string into three block types:
	\begin{itemize}[leftmargin=*]
		\item \textbf{Literal ($L$):} Emits a single character coded via an alphabet range model (or escaped to a side stream).
		\item \textbf{Dictionary Reference ($D$):} Emits a triplet $(\text{word\_rank}, \text{offset}, \text{length})$ referencing a factor contained entirely within a single dictionary entry $w \in \Sdict$.
		\item \textbf{Self-Reference ($S$):} Emits a pair $(\text{distance}, \text{length})$ copying from $s[1..i-1]$. To prevent representation overhead, $S$ blocks enforce $\text{length} \ge 3$. Overlapping copies ($\text{distance} < \text{length}$) are permitted as in standard LZ77.
	\end{itemize}
	
	\paragraph*{Bit-optimal Viterbi parser.}
	Let $n = |s|$. The optimal parse is computed via dynamic programming over an $(n+1) \times 3$ state grid, where state $t \in \{L, D, S\}$ records the block type ending at position $i$:
	\begin{equation}
		\mathrm{dp}[i][t] = \min_{t' \in \{L, D, S\}, \, \text{block } b \text{ ending at } i} \Bigl\{ \mathrm{dp}[i-|b|][t'] - \log_2 P(t \mid t') + \mathrm{Cost}(b) \Bigr\},
	\end{equation}
	where $\mathrm{Cost}(b)$ represents the exact bit cost of encoding block $b$'s parameters under the active probability tables. 
	
	At position $i$, candidate blocks are restricted to:
	\begin{enumerate}[leftmargin=*]
		\item The single literal symbol $s[i]$.
		\item Dictionary factors found by matching $s[i..n]$ against a Directed Acyclic Word Graph (DAWG) or Suffix Automaton (SAM) constructed over $\Sdict$.
		\item Self-reference matches found via a sliding 3-gram index over $s[1..i-1]$, restricted to the $12$ most recent occurrences and capped at length $256$.
	\end{enumerate}
	
	\paragraph*{Context initialization and Viterbi training.}
	Probability models are initialized during context construction. An iterative Viterbi training loop performs $R=3$ refinement passes over $D_{\mathrm{train}}$:
	\begin{enumerate}[leftmargin=*]
		\item \textbf{Pass 0:} Parse $D_{\mathrm{train}}$ using uniform heuristic block costs to gather initial transition and parameter distributions.
		\item \textbf{Passes 1--3:} Re-parse $D_{\mathrm{train}}$ using the dynamic programming parser driven by logarithmic probabilities estimated from the preceding pass, applying Laplace add-one smoothing.
	\end{enumerate}
	This iterative alignment adjusts transition probabilities $P(t \mid t')$ to reflect true coding overhead, preventing literal over-penalization.
	
	\subsection{Entropy Coder, Escapes, and Stream Format}
	\label{sec:arith}
	
	\paragraph*{Range coding and side stream.}
	All structured block fields are encoded into a unified bitstream using a $32$-bit integer range coder with static cumulative frequency tables. Integer parameter values $v \le 64$ present in the model support are coded directly by the range coder. Out-of-support or large integers ($v > 64$) trigger an arithmetic \texttt{ESCAPE} symbol, and their value $v$ is written to a secondary \textbf{side stream} using Elias-gamma variable-length coding.
	
	\paragraph*{Robustness and unseen symbols.}
	If a character in a test file is absent from the context alphabet learned during training, the literal model emits \texttt{ESCAPE\_CHAR}, followed by the character's full Unicode code point in the side stream. This guarantees complete round-trip reliability under out-of-distribution character drift without requiring dictionary retraining.
	
	\paragraph*{Header and payload encapsulation.}
	The shared context (dictionary $\Sdict$, alphabet, and cumulative probability tables) is serialized into a single zlib-compressed header prefix. Each encoded file body consists of three variable-length header fields (uncompressed string length, arithmetic stream byte count, side stream byte count) followed by the concatenated arithmetic and side streams.
	
	\subsection{Computational Complexity}
	\label{sec:complexity_breakdown}
	
	The computational bounds of the implementation are structured as follows:
	\begin{itemize}[leftmargin=*]
		\item \textbf{Offline Dictionary Selection (MDL):} Candidate generation and seed parsing require $O(N_{\mathrm{train}} \cdot \bar{n}^2)$ time, where $\bar{n}$ is the mean training record length. Backward pruning with three-level incremental evaluation re-parses affected documents in $O(|\Sdict_0| \cdot N_{\mathrm{affected}} \cdot \bar{n})$ amortized time. Checkpoint selection requires $K=25$ full Viterbi parses of $D_{\mathrm{train}}$.
		\item \textbf{Context Training ($R=3$ passes):} Building the static dictionary SAM takes $O(\sum_{w \in \Sdict} |w|)$ time and space. Each Viterbi refinement pass over $D_{\mathrm{train}}$ takes $O(N_{\mathrm{train}} \cdot \bar{n})$ time due to bounded lookup candidates.
		\item \textbf{Per-File Encoding and Decoding:} For a target string $s$ of length $N$:
		\begin{itemize}
			\item Candidate generation is bounded by $O(1)$ operations per position thanks to the SAM dictionary traversal and capped 3-gram lookback ($12$ candidates, max length $256$).
			\item The Viterbi DP parser executes in $O(N)$ time and $O(N)$ space.
			\item Single-pass range encoding/decoding operates in $O(N + K_D \cdot |\Sdict|)$ time, where $K_D$ is the count of dictionary blocks (accounting for MTF rank updates).
		\end{itemize}
	\end{itemize}

\subsection{Experimental Evaluation: a Dual-Source Codec on Short Records}
\label{sec:experiments}

This section evaluates the practical dual-source codec of Section~\ref{sec:codec}, not the combinatorial measure itself. The codec uses a bit-cost-guided parser, may emit overlapping copies, and bounds its candidate set, so the phrase-count results of Sections~\ref{sec:theory} and~\ref{sec:algorithm} do not apply to it verbatim. All compressors are evaluated in the same setting: each record is compressed independently and must be decodable on its own from a shared header or dictionary, which preserves random access. Solid-archive compressors that exploit redundancy across records are therefore out of scope. The experiments address four questions:
\begin{enumerate}[leftmargin=*]
	\item \textbf{Size.} How much smaller is the codec than general-purpose and dictionary-based compressors when the baselines are given tuned dictionaries and minimal framing?
	\item \textbf{Components.} What do the dictionary, the self-reference, the MDL selection and the pruning mode each contribute?
	\item \textbf{Trained statistics.} Is the gain explained by the dictionary-based parse, or would a simple context model trained on the same data do as well?
	\item \textbf{Amortization.} How does the comparison depend on the number of records over which the dictionary or header cost is spread?
\end{enumerate}
The comparison is deliberately conservative: baselines are tuned on held-out validation data, they are given dictionary budgets from 512\,B to 32\,KiB, including budgets matched to the size of the codec's dictionary and header, and we report the cases in which the codec does not win as well as those in which it does.

\subsection{Corpora and Experimental Protocol}
\label{sec:protocol}

\paragraph*{Corpora.} Two synthetic and two real corpora are used.
\begin{itemize}[leftmargin=*, itemsep=1pt]
	\item \textbf{Synthetic genomic.} Each record is a fixed 52-character template followed by a tandem repeat of one motif drawn from a pool of 100 motifs of 3 to 6 bases, repeated between 4 and 100 times (mean length $\approx 288$\,B). Tandem repeats are present by construction.
	\item \textbf{Synthetic JSON logs.} Near-identical event lines varying in the minute and second of a timestamp, a server name (5 values), an HTTP-like status (7-entry skewed list), an event name (5 values) and a session identifier ($10^6$ values) (mean length $\approx 114$\,B).
	\item \textbf{Linux logs.} Rows of the Loghub \texttt{Linux\_2k} structured log, each converted into a JSON object whose keys are the column names (mean length $\approx 191$\,B).
	\item \textbf{Real genomic.} About one thousand genomic records truncated to 200\,B (mean length $\approx 200$\,B).
\end{itemize}
Because the synthetic generators are known, the Shannon entropy of the source can be computed exactly: $13.1$ bits ($1.64$\,B) per record for the genomic generator and $38.0$ bits ($4.76$\,B) per record for the JSON generator. These figures are floors for the per-record payload of any decodable code and do not include the cost of describing a dictionary.

\paragraph*{Splits and draws.} For the synthetic corpora, training and test records are generated with independent seeds ($s$ and $s+10^4$); the validation set (100 records) is the continuation of the training stream and is used only to select baseline hyper-parameters. We use $N_{\mathrm{train}}=200$ and generate $N_{\mathrm{test}}=1{,}000$ test records for each synthetic corpus, from which those identical to a training record are removed (this concerns only the synthetic genomic generator, whose records are drawn from a template, a pool of 100 motifs and a repetition count; $979$ test records remain on average, none are removed for JSON). For the real corpora, records are partitioned at random, with quotas scaled down proportionally when the corpus is small (about 770 test records for real genomic data, 1{,}000 for the Linux logs). In all cases $\lambda=N_{\mathrm{test}}/N_{\mathrm{train}}=5$. Every figure is a mean over 3 independent draws.

\paragraph*{Methods compared.}
\begin{itemize}[leftmargin=*, itemsep=1pt]
	\item \textbf{Codec.} The dual-source codec with MDL dictionary selection (Algorithm~\ref{alg:mdl}), with exact backward pruning ($m=-1$) as the reference and lazy pruning ($m=1$) as a variant, checkpoint selection by the true codec cost ($K=25$ checkpoints), computed with the raw size of the header so that the selection criterion and the evaluation charge the same quantity, and three Viterbi re-estimation passes. The header (alphabet, dictionary $\Sdict$ and probability tables) is included in every size.
	\item \textbf{General-purpose compressors (no dictionary).} zlib (level~6 with its container; level~9 as raw DEFLATE), zstd level~19, Brotli quality~11, LZMA (raw LZMA1) and bzip2. Container overhead is removed wherever the format allows it: zstd is run without magic number, checksum, content size or dictionary identifier.
	\item \textbf{Dictionary-based compressors.} zlib-9 (raw DEFLATE with a preset dictionary) and zstd-19 with a dictionary, each with three dictionary sources: training records sampled at random (``sampled''), the dictionary produced by zstd's trainer (``trained''), and the content of the MDL dictionary $\Sdict$ used as raw dictionary text.
	\item \textbf{Idealized controls.} Static order-$k$ context models ($k=1,2,3$, Witten--Bell interpolation) trained on the same training records. Their cost is the ideal code length of each record plus one byte for arithmetic-coder termination, and the model is charged at the size of its raw serialization. They are not implemented as decoders; they measure what trained statistics alone can achieve. The codec's literal model, by contrast, does not condition on the preceding symbols.
	\item \textbf{Optimistic bounds.} Brotli, LZMA and PPMd offer no custom-dictionary interface in the Python bindings we use. Their dictionary-primed cost is estimated from the marginal cost of a record appended to the dictionary. This estimator is biased in favor of the baseline (Appendix), so these rows are reported there only and are excluded from all comparisons.
\end{itemize}

\paragraph*{Accounting and tuning.} The total size of a method is the sum of the compressed record sizes (payload), of one length field per record (the same variable-length integer for all methods, so that every record can be located), and of the shared dictionary or header. For each dictionary-based family the dictionary size is swept from 512\,B to 32\,KiB, with two additional matched budgets equal to the size of $\Sdict$ and to the size of the codec header; the budget that minimizes the expected total size on the validation set is retained. The codec needs no budget, since the MDL criterion chooses the dictionary size. Every method is charged for its dictionary, header or model at its \emph{raw} (uncompressed) size: for the codec, the whole uncompressed header (alphabet, dictionary $\Sdict$ and probability tables, not only the words of $\Sdict$); for the baselines, the raw dictionary text; for the context controls, the raw serialized model. This is the rule least favorable to the codec, whose header is stored compressed ($1.8$ to $3.1$ times smaller than its raw size); the appendix tables report both sizes, and Section~\ref{sec:main_results} gives the effect of charging stored sizes. Baseline budgets were selected on the validation set under the raw rule. Every decodable method is decoded and compared byte for byte with its input, and the codec's decoder is rebuilt from the serialized header alone.

\paragraph*{Reading the tables.} $\Delta=1-\text{codec}/\text{method}$ is the relative gain of the codec with respect to a method, averaged over draws of the per-draw gain. A positive $\Delta$ means that the codec is smaller than the method by $\Delta\%$ of the method's size; a negative $\Delta$ means that the method is smaller than the codec.

\subsection{Main Results}
\label{sec:main_results}

\begin{table}[t]
\centering
\footnotesize
\setlength{\tabcolsep}{3.2pt}
\caption{Scenario~A: total compressed size (B) of the held-out records and gain of the codec against each baseline.}
\label{tab:main}
\resizebox{\linewidth}{!}{%
\begin{tabular}{lrrrrrrrr}
\toprule
Method & \multicolumn{2}{c}{Synth.\ genomic} & \multicolumn{2}{c}{Synth.\ JSON} & \multicolumn{2}{c}{Linux logs} & \multicolumn{2}{c}{Real genomic} \\
\cmidrule(lr){2-3}\cmidrule(lr){4-5}\cmidrule(lr){6-7}\cmidrule(lr){8-9}
 & Total (B) & $\Delta$ (\%) & Total (B) & $\Delta$ (\%) & Total (B) & $\Delta$ (\%) & Total (B) & $\Delta$ (\%) \\
\midrule
\multicolumn{9}{l}{\emph{Ours}} \\
\textbf{Codec} (MDL, exact pruning) & 10\,847 & ref. & 17\,566 & ref. & 32\,115 & ref. & 68\,063 & ref. \\
Codec, lazy pruning ($m{=}1$) & 11\,304 & +4.0 & 18\,544 & +5.3 & 32\,255 & +0.4 & 74\,656 & +8.8 \\
\midrule
\multicolumn{9}{l}{\emph{Dictionary-based baselines (decodable)}} \\
zlib-9 + trained dict. & 10\,877 & +0.3 & 22\,521 & +22.0 & 37\,170 & +13.6 & 73\,874 & +7.9 \\
zlib-9 + sampled dict. & 11\,901 & +8.8 & \textbf{22\,336} & +21.4 & 40\,485 & +20.5 & 74\,835 & +9.0 \\
zlib-9 + MDL dict.\ $S$ & \textbf{10\,844} & 0.0 & 22\,882 & +23.2 & \textbf{36\,393} & +11.8 & 74\,333 & +8.4 \\
zstd-19 + trained dict. & 19\,754 & +45.1 & 31\,831 & +44.8 & 45\,402 & +29.3 & \textbf{73\,295} & +7.1 \\
zstd-19 + sampled dict. & 20\,738 & +47.7 & 31\,830 & +44.8 & 51\,081 & +37.0 & 86\,139 & +21.0 \\
zstd-19 + MDL dict.\ $S$ & 20\,244 & +46.4 & 33\,003 & +46.8 & 47\,772 & +32.8 & 86\,903 & +21.7 \\
\midrule
\multicolumn{9}{l}{\emph{General-purpose baselines (no dictionary)}} \\
Brotli-11 & 32\,128 & +66.2 & 105\,746 & +83.4 & 145\,976 & +78.0 & 103\,859 & +34.5 \\
zstd-19 & 36\,451 & +70.2 & 111\,007 & +84.2 & 154\,182 & +79.2 & 110\,594 & +38.5 \\
zlib-9 (raw deflate) & 41\,792 & +74.0 & 108\,863 & +83.9 & 148\,653 & +78.4 & 110\,199 & +38.2 \\
zlib-6 (framed) & 47\,666 & +77.2 & 114\,863 & +84.7 & 154\,785 & +79.2 & 114\,832 & +40.7 \\
LZMA (raw) & 48\,709 & +77.7 & 110\,163 & +84.0 & 155\,187 & +79.3 & 115\,571 & +41.1 \\
bzip2 & 67\,493 & +83.9 & 137\,482 & +87.2 & 183\,323 & +82.5 & 133\,760 & +49.1 \\
\bottomrule
\end{tabular}}
\par\vspace{1mm}\begin{minipage}{\linewidth}\footnotesize $\Delta=1-\text{codec}/\text{method}$, mean over the independent draws of the per-draw gain: positive means the codec is smaller, negative that the method is smaller. Bold: best decodable dictionary-based baseline. Totals include the dictionary or header and the record index. Every method is charged for its dictionary, header or model at its \emph{raw} (uncompressed) size: for the codec, the whole uncompressed header (alphabet, dictionary $S$ and probability tables); for the baselines, the raw dictionary text; for the context controls, the raw serialized model. Idealized controls are reported in Table~\ref{tab:ablation}; optimistic bounds in the appendix.\end{minipage}
\end{table}

\paragraph*{Against general-purpose compressors.} Without a dictionary, independent compression of short records is ineffective. Brotli-11 is the best general-purpose baseline on all four corpora, with ratios of $11.4\%$ (synthetic genomic), $92.7\%$ (JSON), $76.47\%$ (Linux) and $67.44\%$ (real genomic); the others are worse, and on the JSON records all general-purpose outputs lie between $93\%$ and $121\%$ of the raw size. The codec reaches ratios of $3.8\%$, $15.4\%$, $16.8\%$ and $44.2\%$, that is $34.5$--$83.4\%$ smaller than Brotli-11 ($\Delta=+66.2$, $+83.4$, $+78.0$ and $+34.5\%$).

\paragraph*{Against dictionary-based compressors.} Equipped with tuned dictionaries, zlib and zstd become much stronger, yet the codec remains smaller than every dictionary-based variant on three corpora and ties with the best one on the fourth. Against zstd-19, the best variant is the trained dictionary on synthetic genomic data ($19{,}754$\,B, $\Delta=+45.1\%$), on the Linux logs ($45{,}402$\,B, $\Delta=+29.3\%$) and on real genomic data ($73{,}295$\,B, $\Delta=+7.1\%$), and the sampled dictionary on JSON ($31{,}830$\,B, $\Delta=+44.8\%$, the trained one being $1$\,B larger). Against zlib-9, the codec is smaller than the best variant on JSON ($17{,}566$ against $22{,}336$\,B with a sampled dictionary, $\Delta=+21.4\%$), on the Linux logs ($32{,}115$ against $36{,}393$\,B with $\Sdict$, $\Delta=+11.8\%$; $37{,}170$\,B with a trained dictionary, $\Delta=+13.6\%$) and on real genomic data ($68{,}063$ against $73{,}874$\,B with a trained dictionary, $\Delta=+7.9\%$; $\Delta=+8.4\%$ and $+9.0\%$ for the MDL and sampled dictionaries). On synthetic genomic data the codec ($10{,}847$\,B) and the two best zlib-9 variants ($10{,}844$\,B with $\Sdict$, $10{,}877$\,B trained) lie within $0.3\%$ of each other, which with three draws we regard as a tie ($\Delta=0.0\%$ and $+0.3\%$); the sampled variant is $8.8\%$ larger than the codec.

\paragraph*{At equal dictionary content.} Giving zlib-9 the MDL dictionary $\Sdict$ isolates the parse and the entropy coder from the dictionary content. The codec is then smaller on JSON ($17{,}566$ against $22{,}882$\,B, $\Delta=+23.2\%$), on the Linux logs ($32{,}115$ against $36{,}393$\,B, $+11.8\%$) and on real genomic data ($68{,}063$ against $74{,}333$\,B, $+8.4\%$), where its payload is $28.8\%$, $16.9\%$ and $14.1\%$ smaller. On synthetic genomic data the payload gain is only $2.4\%$ ($8{,}964$ against $9{,}183$\,B), which DEFLATE copies natively on tandem repeats, and it is cancelled by the larger header of the codec ($904$ against $683$\,B raw, because of its alphabet and probability tables): the totals are equal ($\Delta=0.0\%$). The comparison is therefore not exactly at equal cost, since the codec also pays for its tables. The MDL dictionary is a competitive dictionary text for DEFLATE: it is the best zlib-9 dictionary on synthetic genomic data and on the Linux logs, $2.4\%$ worse than a sampled dictionary on JSON and $0.6\%$ worse than a trained one on real genomic data.

\paragraph*{Sensitivity to the charging rule.} Charging every method at its stored (zlib-compressed) size instead, computed from the totals of the appendix tables without re-tuning the baseline budgets, the codec totals become $10{,}231$, $16{,}403$, $28{,}425$ and $63{,}716$\,B. The ranking against the dictionary-based baselines is unchanged: the codec remains smaller than the best zlib-9 variant on JSON ($\Delta=+21.7\%$), Linux ($+12.0\%$) and real genomic data ($+8.8\%$), and on synthetic genomic data it is $0.2\%$ smaller than zlib-9 with $\Sdict$ ($10{,}249$\,B), still a tie. The rule matters only for the idealized controls: the margin of the codec over the order-2 context model on synthetic JSON falls from $4.1\%$ to $1.6\%$, and the codec remains larger on real genomic data ($50{,}998$\,B for order~2, $\Delta=-24.9\%$). The raw rule used throughout is the conservative one for the codec, whose header compresses $1.8$ to $3.1$ times.

\subsection{Ablations and Idealized Controls}
\label{sec:ablation}

\begin{table}[t]
\centering
\footnotesize
\setlength{\tabcolsep}{3.2pt}
\caption{Ablations of the codec and idealized controls (Scenario~A). Same conventions as Table~\ref{tab:main}; for the controls the model size is charged in the Total column (raw size).}
\label{tab:ablation}
\resizebox{\linewidth}{!}{%
\begin{tabular}{lrrrrrrrr}
\toprule
Method & \multicolumn{2}{c}{Synth.\ genomic} & \multicolumn{2}{c}{Synth.\ JSON} & \multicolumn{2}{c}{Linux logs} & \multicolumn{2}{c}{Real genomic} \\
\cmidrule(lr){2-3}\cmidrule(lr){4-5}\cmidrule(lr){6-7}\cmidrule(lr){8-9}
 & Total (B) & $\Delta$ (\%) & Total (B) & $\Delta$ (\%) & Total (B) & $\Delta$ (\%) & Total (B) & $\Delta$ (\%) \\
\midrule
\multicolumn{9}{l}{\emph{Codec variants}} \\
\textbf{Codec} (MDL, exact pruning) & 10\,847 & ref. & 17\,566 & ref. & 32\,115 & ref. & 68\,063 & ref. \\
Codec, lazy pruning ($m{=}1$) & 11\,304 & +4.0 & 18\,544 & +5.3 & 32\,255 & +0.4 & 74\,656 & +8.8 \\
Codec w/o self-reference (RLZ) & 60\,355 & +82.0 & 17\,565 & 0.0 & 32\,305 & +0.6 & 66\,157 & $-$2.9 \\
Codec w/o dictionary ($S{=}\emptyset$) & 23\,263 & +53.4 & 74\,122 & +76.3 & 112\,225 & +71.4 & 82\,641 & +17.6 \\
Codec, $S$ = chunks of zstd-trained dict. & 11\,623 & +6.7 & 18\,456 & +4.8 & 38\,539 & +16.7 & 63\,198 & $-$7.7 \\
\midrule
\multicolumn{9}{l}{\emph{Idealized trained context models (not decoded; ideal code length $+1$\,B per record)}} \\
Order-1 context model (ideal) & 71\,383 & +84.8 & 30\,389 & +42.2 & 67\,052 & +52.1 & 55\,129 & $-$23.5 \\
Order-2 context model (ideal) & 71\,627 & +84.8 & 18\,314 & +4.1 & 42\,269 & +24.0 & 54\,077 & $-$25.9 \\
Order-3 context model (ideal) & 70\,715 & +84.7 & 19\,538 & +10.1 & 42\,829 & +25.0 & 60\,604 & $-$12.3 \\
\bottomrule
\end{tabular}}
\end{table}

\paragraph*{The dictionary dominates on short records.} Removing the dictionary ($S=\emptyset$, the codec then behaves as an adaptive LZ77 coder) multiplies the total size by $2.14$ (synthetic genomic, $23{,}263$\,B), $4.22$ (JSON, $74{,}122$\,B), $3.49$ (Linux, $112{,}225$\,B) and $1.21$ (real genomic, $82{,}641$\,B). The external dictionary is thus the primary engine of compression, least so on the real genomic data.

\paragraph*{Self-reference is decisive on tandem repeats and irrelevant elsewhere.} Removing dynamic self-reference (pure RLZ with the same parser) multiplies the size by $5.56$ on synthetic genomic data ($60{,}355$ against $10{,}847$\,B), a corpus built from tandem repeats: there the dictionary alone is \emph{worse} than no dictionary ($23{,}263$\,B for $S=\emptyset$), and the combination of both sources is what makes the codec effective. This is consistent with the repetition bound of Proposition~\ref{prop:idem-adapt}, under which a motif repeated $k$ times is covered by one dictionary block plus about $\log_2 k$ prefix blocks in the non-overlapping model; the codec, which allows overlapping copies, needs a single self-reference block whatever $k$, so its gain is, if anything, larger than that model predicts. On the three other corpora the effect is small: $0.0\%$ on JSON, $+0.6\%$ on the Linux logs, and $-2.9\%$ on real genomic data, where RLZ is \emph{smaller} ($66{,}157$ against $68{,}063$\,B). Their records contain little internal repetition beyond what the dictionary covers. Self-reference thus pays only when records repeat their own content.

\paragraph*{MDL selection and pruning mode.} Exact backward pruning ($m=-1$) is better than or equal to lazy pruning ($m=1$) on all four corpora: $10{,}847$ against $11{,}304$\,B on synthetic genomic data ($\Delta=+4.0\%$), $17{,}566$ against $18{,}544$\,B on JSON ($+5.3\%$), $32{,}115$ against $32{,}255$\,B on Linux ($+0.4\%$, within the noise) and $68{,}063$ against $74{,}656$\,B on real genomic data ($+8.8\%$). Lazy pruning always yields the larger raw header ($1{,}628$ against $904$\,B, $2{,}838$ against $2{,}019$\,B, $6{,}758$ against $5{,}934$\,B and $15{,}816$ against $9{,}495$\,B), and a smaller payload only on synthetic genomic data and on the Linux logs ($8{,}697$ against $8{,}964$\,B and $24{,}497$ against $25{,}181$\,B): the words it keeps do not save enough payload to pay for their raw description. Replacing the MDL dictionary by chunks of a zstd-trained dictionary gives a larger codec on the synthetic and Linux corpora ($\Delta=+6.7\%$, $+4.8\%$ and $+16.7\%$) but a $7.1\%$ \emph{smaller} one on real genomic data ($63{,}198$ against $68{,}063$\,B, $\Delta=-7.7\%$, with both a smaller header and a smaller payload), so the MDL selection is not optimal there. The dictionaries selected here are much smaller than those obtained when the selection criterion charged the stored instead of the raw header (Section~\ref{sec:threats}).

\paragraph*{Trained statistics.} On synthetic genomic data, the context models are far worse than the codec (order~3: $70{,}715$\,B, $\Delta=+84.7\%$), as low-order Markov contexts cannot span variable-length tandem repeats. On the Linux logs, the codec is smaller than the order-2 model ($32{,}115$ against $42{,}269$\,B, $\Delta=+24.0\%$) and the order-3 model ($\Delta=+25.0\%$). On synthetic JSON, it is smaller than the order-2 model ($17{,}566$ against $18{,}314$\,B, $\Delta=+4.1\%$) and the order-3 model ($19{,}538$\,B, $\Delta=+10.1\%$) at $n_{\mathrm{test}}$, although the order-2 model overtakes it beyond $1{,}528$ records (Section~\ref{sec:amortization}). On real genomic data, all three context models are smaller than the codec: order~2 reaches $54{,}077$\,B ($\Delta=-25.9\%$), order~1 $55{,}129$\,B ($\Delta=-23.5\%$) and order~3 $60{,}604$\,B ($\Delta=-12.3\%$), the last being penalized by its $18{,}829$\,B model. Because the codec's literal model does not condition on preceding symbols, records with few long dictionary matches are coded less efficiently than by Markovian literal models. These controls are idealized (ideal code length, not decoded), so the gap to a real decoder would be somewhat smaller.

\paragraph*{Distance to the entropy floor.} On the synthetic corpora, the codec's payload (without the length field) is about $9.2$\,B per record on the genomic corpus (approx.\ $5.6$ times the $1.64$\,B source entropy) and about $14.5$\,B per record on JSON (approx.\ $3.1$ times the $4.76$\,B source entropy). The remaining gap reflects the cost of describing the source through a discrete dictionary and token-level bit coding.

\subsection{Amortization of the Dictionary}
\label{sec:amortization}

\begin{figure}[htbp]
	\centering
	\begin{minipage}{0.48\linewidth}
		\centering
		\includegraphics[width=\linewidth]{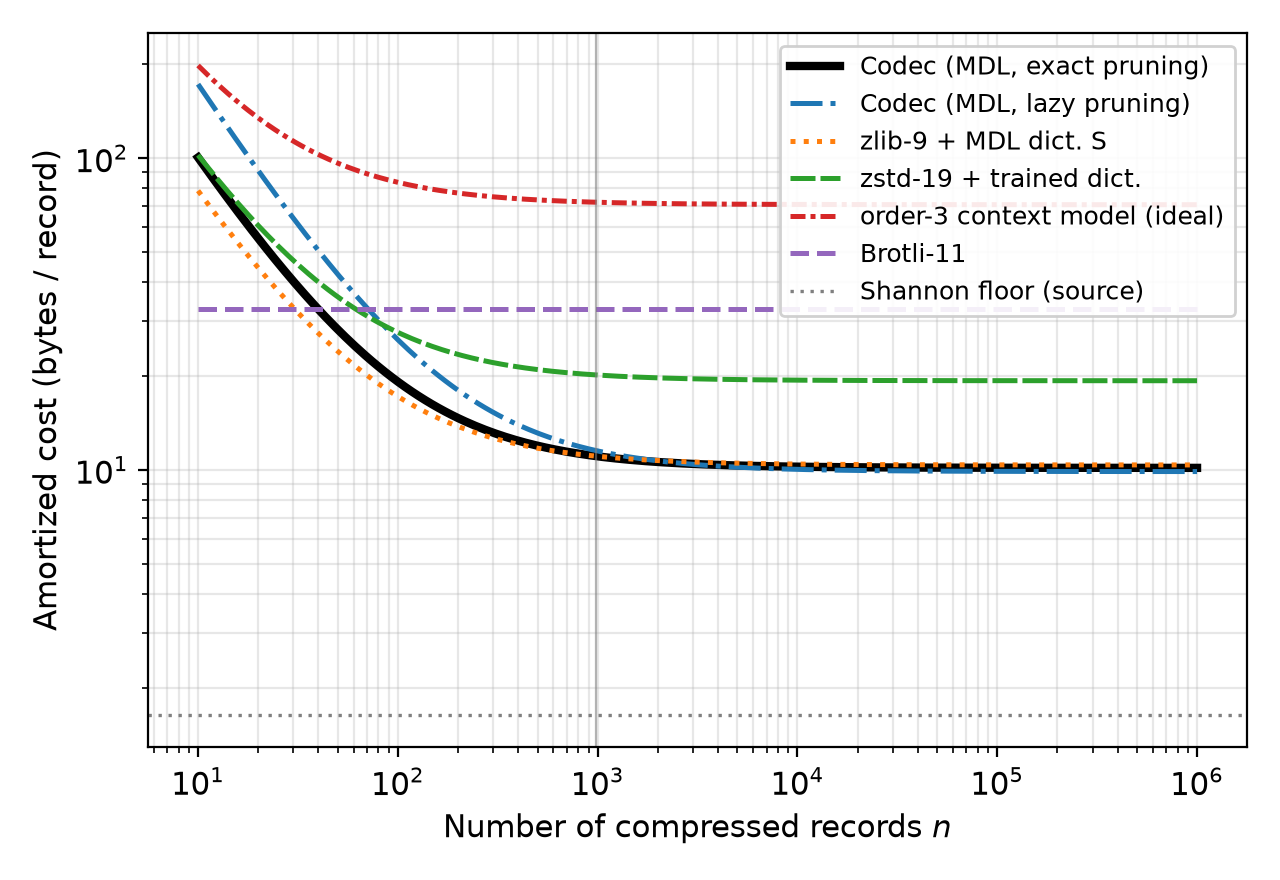}\\[-1mm]
		{\scriptsize (a) Synthetic genomic}
	\end{minipage}\hfill
	\begin{minipage}{0.48\linewidth}
		\centering
		\includegraphics[width=\linewidth]{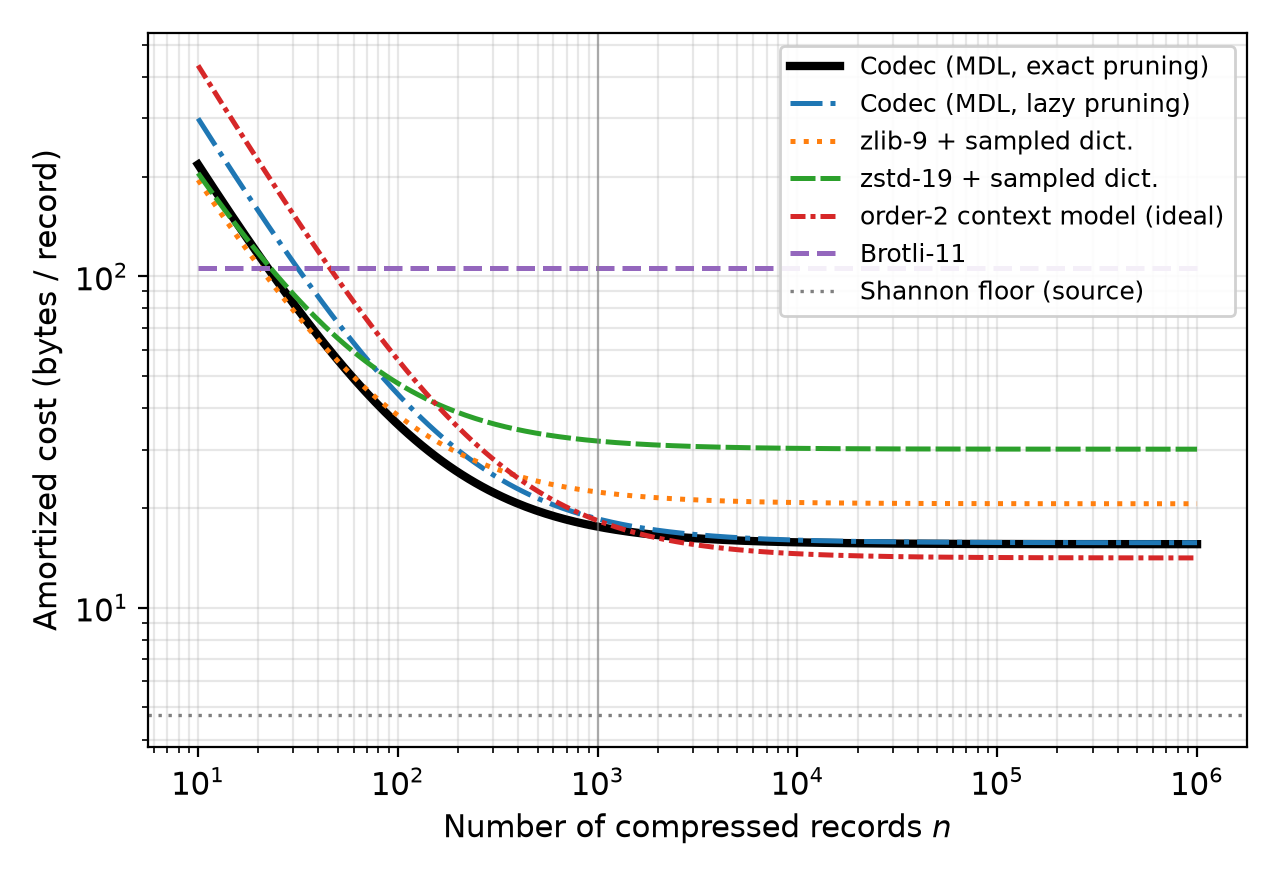}\\[-1mm]
		{\scriptsize (b) Synthetic JSON}
	\end{minipage}
	
	\vspace{3mm}
	
	\begin{minipage}{0.48\linewidth}
		\centering
		\includegraphics[width=\linewidth]{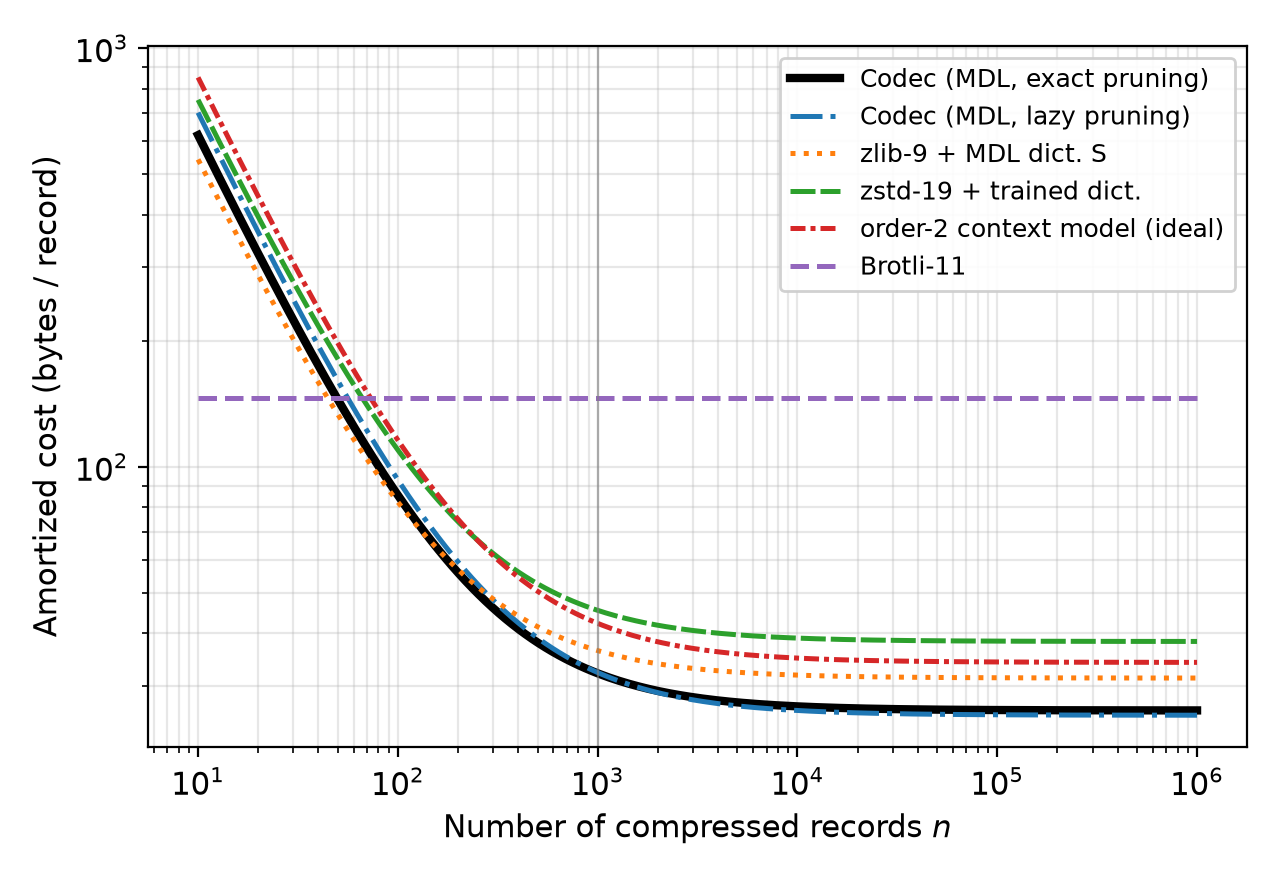}\\[-1mm]
		{\scriptsize (c) Linux logs}
	\end{minipage}\hfill
	\begin{minipage}{0.48\linewidth}
		\centering
		\includegraphics[width=\linewidth]{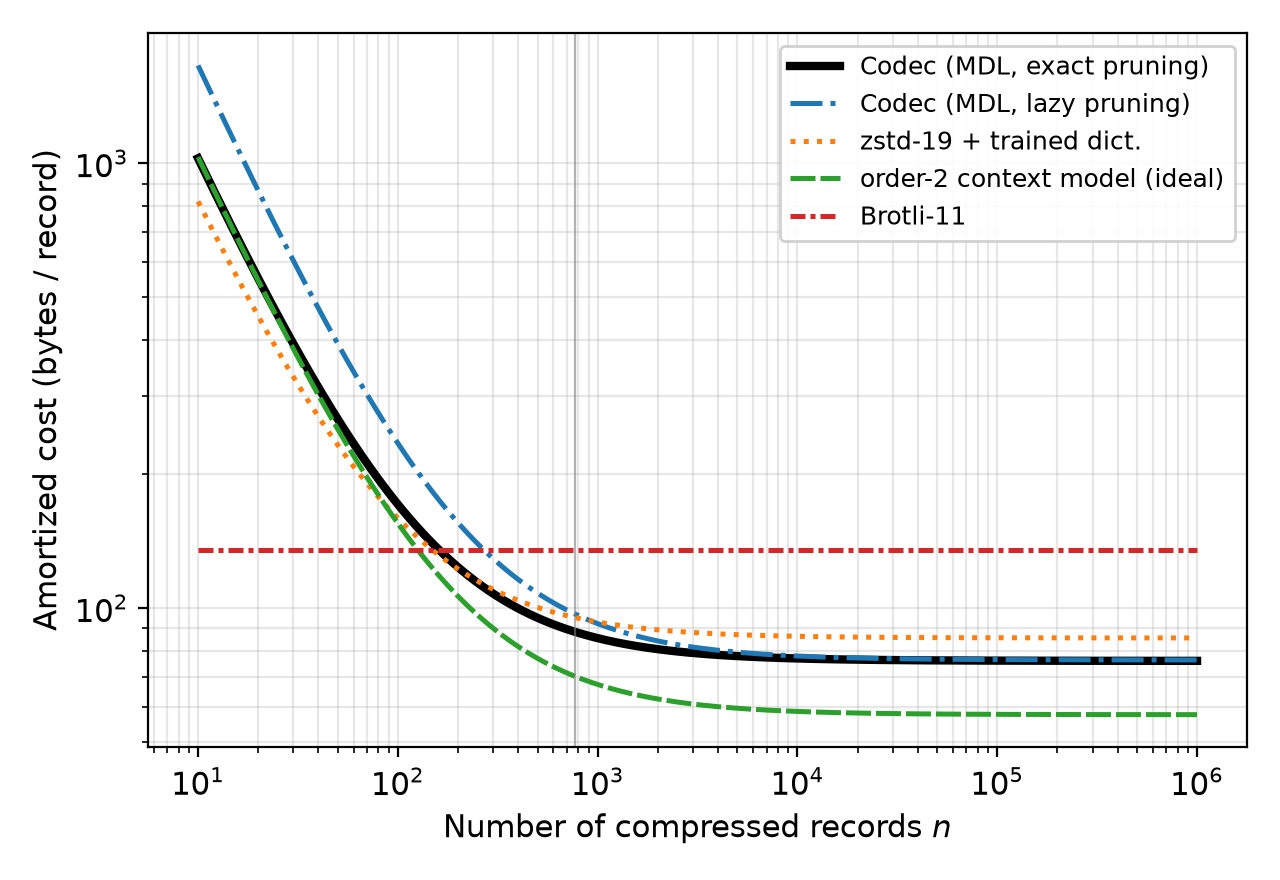}\\[-1mm]
		{\scriptsize (d) Real genomic}
	\end{minipage}
	
	\caption{Amortized cost per record (bytes/record) as a function of the number of compressed records $n$, comparing the dual-source codec (thick curve) and its lazy variant against the best dictionary-based baseline, the best idealized control, Brotli-11 and, for the synthetic corpora, the Shannon floor (baseline and control selected at $n=n_{\mathrm{test}}$, vertical line).}
	\label{fig:nsweep}
\end{figure}

The sizes above are totals over about $10^3$ records, over which the codec header and the baseline dictionaries are amortized. Table~\ref{tab:nsweep} and Figure~\ref{fig:nsweep} give the cost per record as a function of the number $n$ of records, modelled as $(\text{dictionary})/n+\bar c$ where $\bar c$ is the measured mean per-record cost (payload and length field). For $n$ beyond the size of the test set this extrapolates the measured per-record cost with a frozen dictionary; the dictionary term of every method is charged as in Table~\ref{tab:main} (raw size).

\begin{table}[t]
\centering
\footnotesize
\setlength{\tabcolsep}{3.5pt}
\caption{Amortized cost per record (B/record: payload and index, plus dictionary or header divided by $n$) as a function of the number $n$ of records, for the codec, the best dictionary-based baseline and the best idealized control of each corpus (selected at $n=n_{\mathrm{test}}$, the size of the test set: 979, 1\,000, 1\,000, 770 records for the four corpora in order). The last column gives the range of $n$ over which the codec is the smaller one. Values for $n>n_{\mathrm{test}}$ extrapolate the measured per-record cost with a frozen dictionary. Every method is charged for its dictionary, header or model at its \emph{raw} (uncompressed) size: for the codec, the whole uncompressed header (alphabet, dictionary $S$ and probability tables); for the baselines, the raw dictionary text; for the context controls, the raw serialized model.}
\label{tab:nsweep}
\resizebox{\linewidth}{!}{%
\begin{tabular}{llrrrrrrl}
\toprule
Corpus & Method & Dict.\ (B) & $n{=}10$ & $n{=}10^2$ & $n{=}n_{\mathrm{test}}$ & $n{=}10^4$ & $n\to\infty$ & Codec smaller for \\
\midrule
Synth.\ genomic & \textbf{Codec} (MDL, exact pruning) & 904 & 100.5 & 19.2 & 11.1 & 10.2 & 10.16 & -- \\
 & zlib-9 + MDL dict.\ $S$ & 683 & 78.6 & 17.2 & 11.1 & 10.4 & 10.38 & $n > 989$ \\
 & Order-3 context model (ideal) & 1\,274 & 198.3 & 83.7 & 72.2 & 71.0 & 70.92 & all $n$ \\
\midrule
Synth.\ JSON & \textbf{Codec} (MDL, exact pruning) & 2\,019 & 217.4 & 35.7 & 17.6 & 15.7 & 15.55 & -- \\
 & zlib-9 + sampled dict. & 1\,747 & 195.3 & 38.1 & 22.3 & 20.8 & 20.59 & $n > 54$ \\
 & Order-2 context model (ideal) & 4\,186 & 432.7 & 56.0 & 18.3 & 14.5 & 14.13 & $n < 1\,528$ \\
\midrule
Linux logs & \textbf{Codec} (MDL, exact pruning) & 5\,934 & 619.6 & 85.5 & 32.1 & 26.8 & 26.18 & -- \\
 & zlib-9 + MDL dict.\ $S$ & 5\,109 & 542.1 & 82.4 & 36.4 & 31.8 & 31.29 & $n > 162$ \\
 & Order-2 context model (ideal) & 8\,170 & 851.1 & 115.8 & 42.3 & 34.9 & 34.10 & all $n$ \\
\midrule
Real genomic & \textbf{Codec} (MDL, exact pruning) & 9\,495 & 1025.5 & 171.0 & 88.4 & 77.0 & 76.06 & -- \\
 & zstd-19 + trained dict. & 7\,348 & 820.5 & 159.1 & 95.2 & 86.4 & 85.64 & $n > 224$ \\
 & Order-2 context model (ideal) & 9\,743 & 1031.9 & 155.0 & 70.2 & 58.5 & 57.58 & $n < 13$ \\
\bottomrule
\end{tabular}}
\end{table}

\paragraph*{Against dictionary-based compressors.} The raw header of the codec is only $1.2$ to $1.3$ times larger than that of the best dictionary-based baseline ($9{,}495$ against $7{,}348$\,B on real genomic data, $5{,}934$ against $5{,}109$\,B on Linux logs, $2{,}019$ against $1{,}747$\,B on JSON, $904$ against $683$\,B on synthetic genomic data), while its asymptotic per-record cost is lower by $11.2\%$ (real genomic, $76.06$ against $85.64$\,B), $16.3\%$ (Linux), $24.5\%$ (JSON) and $2.1\%$ (synthetic genomic). The codec is smaller than the best baseline for $n>54$ on synthetic JSON (against zlib-9 with a sampled dictionary), $n>162$ on the Linux logs (against zlib-9 with $\Sdict$), $n>224$ on real genomic data (against zstd-19 with a trained dictionary) and $n>989$ on synthetic genomic data (against zlib-9 with $\Sdict$). At $n_{\mathrm{test}}$ it is therefore ahead of the best baseline on JSON, Linux and real genomic data, and level with it on synthetic genomic data ($n_{\mathrm{test}}=979$ against a break-even of $989$), which is the reading of Table~\ref{tab:main}.

\paragraph*{Against trained context models.} The comparison depends on the corpus. On synthetic genomic data, the codec is smaller than the order-3 model for every $n$ (asymptotically $10.16$ against $70.92$\,B per record), and so it is on the Linux logs against the order-2 model (asymptotically $23.2\%$ smaller, $26.18$ against $34.10$\,B). On synthetic JSON, the codec is smaller than the order-2 model only for $n<1{,}528$; beyond, the order-2 model is smaller, asymptotically by $9.1\%$ ($14.13$ against $15.55$\,B per record). On real genomic data, the order-2 model is smaller as soon as $n>13$ (asymptotically $57.58$ against $76.06$\,B), because its state table ($9{,}743$\,B) is of similar size to the codec's header ($9{,}495$\,B) and its per-record cost is $24\%$ lower.

\subsection{Summary of Findings}
\label{sec:summary_results}

\begin{enumerate}[leftmargin=*, itemsep=2pt]
	\item Against the best general-purpose per-record compressor (Brotli-11), the codec is $34.5$--$83.4\%$ smaller on every corpus (ratios $3.8$--$44.2\%$).
	\item Against tuned dictionary baselines, the codec is smaller than the best zstd-19 variant on all four corpora ($\Delta=+7.1\%$ to $+45.1\%$) and than the best zlib-9 variant on JSON ($+21.4\%$), Linux ($+11.8\%$) and real genomic data ($+7.9\%$); on synthetic genomic data it ties with zlib-9 ($\Delta=0.0\%$ to $+0.3\%$). The ranking is the same when every method is charged at its stored instead of its raw size.
	\item At equal dictionary content (zlib-9 with $\Sdict$), the codec is smaller on JSON ($+23.2\%$), Linux logs ($+11.8\%$) and real genomic data ($+8.4\%$), with payloads $14$--$29\%$ smaller; on synthetic genomic data its $2.4\%$ payload gain is cancelled by its larger header ($0.0\%$).
	\item The dictionary accounts for most of the gain (size multiplied by $1.2$--$4.2$ without it). Dynamic self-reference is decisive on the tandem-repeat corpus (size multiplied by $5.6$ without it, and the dictionary alone is worse than no dictionary) and has no measurable effect on the other three corpora.
	\item Exact pruning is better than or equal to lazy pruning on all four corpora, but the MDL dictionary is not optimal on real genomic data, where zstd-trained chunks give a $7.1\%$ smaller codec.
	\item The raw header of the codec is only $1.2$--$1.3$ times that of the best baseline dictionary and is amortized beyond $n=54$ to $989$ records. Idealized trained order-2 context models are smaller than the codec on real genomic data for $n>13$ and on synthetic JSON for $n>1{,}528$, but not on the Linux logs or on synthetic genomic data. The codec is therefore not best-in-class.
\end{enumerate}

\subsection{Threats to Validity and Limitations}
\label{sec:threats}
\begin{itemize}[leftmargin=*, itemsep=2pt]
	\item \textbf{Model vs.\ codec gap.} The empirical codec optimizes encoded bit length by dynamic programming rather than pure phrase count $C_S^{\mathrm{adapt}}(s)$. Its performance reflects downstream entropy-modeling choices, including a context-free literal model and bounded candidate factors (matches of at most $256$ characters).
	\item \textbf{Dictionary accounting sensitivity.} All methods are charged at the raw size of their dictionary, header or model. Charging stored (zlib-compressed) sizes instead changes the margins against the dictionary-based baselines by at most one point and leaves the ranking unchanged (Section~\ref{sec:main_results}); it lowers the margin over the order-2 model on JSON, since the tables of both models compress well ($1.8$ to $3.1$ times for the codec header). Baseline budgets were selected on the validation set under the raw rule only.
	\item \textbf{Statistical uncertainty.} Every figure is a mean over only $3$ draws, no confidence intervals are reported, and differences of a few percent (the tie with zlib-9 on synthetic genomic data, $\Delta=0.0\%$ to $+0.3\%$; $\Delta=+0.4\%$ between the pruning modes and $+0.6\%$ for RLZ on the Linux logs) should not be over-interpreted.
	\item \textbf{Synthetic corpora.} The synthetic generators are the authors'; the structure they contain (tandem repeats, near-identical templates) favors dictionary methods, and the entropy floors apply only to them.
	\item \textbf{Selection criterion and dictionary size.} The MDL checkpoint selection charges the raw header, like the evaluation. In an earlier run in which the criterion used the stored (compressed) header, which under-charges each dictionary word, the selected dictionaries were much larger (raw header of $4{,}642$, $11{,}013$ and $15{,}078$\,B instead of $904$, $2{,}019$ and $5{,}934$\,B on synthetic genomic, JSON and Linux; unchanged at $9{,}495$\,B on real genomic data) and the codec was larger than zlib-9 with dictionaries on the synthetic corpora. The selection criterion must therefore match the accounting rule. On real genomic data, zstd-trained chunks still give a smaller codec, the selection is made on the training records only, and the effect of the deployment parameter $\lambda$ and of a size budget was not studied.
	\item \textbf{Controls and bounds.} Context models represent entropy bounds rather than full decodable codecs. Stronger statistical compressors (PPM or context mixing with trained priors) and Brotli with custom dictionaries were not benchmarked.
	\item \textbf{Corpora scale.} Experiments focus on short-record batches ($10^3$ records, training sets of $150$--$200$ records). Scaling training sets or record lengths may alter the balance between static dictionary matching and statistical context models.
	\item \textbf{Offline cost.} MDL dictionary selection evaluates candidate parses across $K=25$ checkpoints, representing an offline pre-computation cost not charged to online stream encoding.
\end{itemize}

\newpage

	\section{Summary of Core Properties}
	
	\small
	\begin{longtable}{p{4.8cm}p{4.2cm}p{5.1cm}}
		\caption{Core structural properties.}\label{tab:summary}\\
		\toprule
		Property & Non-adaptive $\CS{\cdot}$ & Adaptive $\CadaptSym$ \\
		\midrule
		\endfirsthead
		\toprule
		Property & Non-adaptive $\CS{\cdot}$ & Adaptive $\CadaptSym$ \\
		\midrule
		\endhead
		\bottomrule
		\endlastfoot
		Bounds $1 \le \cdot \le \len{s}$ & \checkmark (Prop.~\ref{prop:bounds}) & \checkmark (Prop.~\ref{prop:bounds}) \\
		Monotonicity in $\Sdict$ & \checkmark (Prop.~\ref{prop:monotonicity}) & \checkmark (Prop.~\ref{prop:monotonicity}) \\
		Monotonicity by contiguous factor & \checkmark (Lem.~\ref{lem:factor-monotone}) & fails (Remark after Lem.~\ref{lem:factor-monotone}) \\
		Concatenation & \checkmark, within $\pm1$ (Prop.~\ref{prop:exact-concat}) & \checkmark, within $\pm1$ with contextual cost (Prop.~\ref{prop:chain-adapt}) \\
		Repetition $w^n$ & within $n-1$, linear in $n$ (Cor.~\ref{cor:powers-exact}) & $\le \Cadapt{w}+\lceil\log_2 n\rceil$, tight (Prop.~\ref{prop:idem-adapt}) \\
		Relation between the two measures & --- & $1+\lceil\log_2\CS{s}\rceil\le\Cadapt{s}\le\CS{s}$, tight (Thm.~\ref{thm:exp}) \\
		Universal floor & $\ge\lceil n/L\rceil$ & $\ge 1+\lceil\log_2\lceil n/L\rceil\rceil$ (Cor.~\ref{cor:floor}) \\
		Counting bound & --- & $\le\binom{n-1}{k-1}(M+n+\sigma)^k$ strings with $\le k$ blocks (Thm.~\ref{thm:combinatorial_counting}) \\
		Separation criterion & --- & all gains $\Delta_j\le0$ $\Rightarrow$ $\Cadapt{s}=\CS{s}$ (Prop.~\ref{prop:adaptation_separation}) \\
		Normalized phrase count & --- & independent of fixed finite $S$ for stationary ergodic sources over $|\Sigma|\ge2$ with $h>0$ (Thm.~\ref{thm:asymptotic_invariance}) \\
		Greedy optimality & \checkmark (Sec.~\ref{sec:rlz-relation}) & \checkmark (Thm.~\ref{thm:greedy-choice}) \\
		Empty dictionary & --- & non-overlapping LZ77 phrase count (Prop.~\ref{prop:grammar-coincide}) \\
		Domination & $=z_{\mathrm{RLZ}}$ (word-segmented) & $\le\min\{z_{\mathrm{RLZ}},z_{\mathrm{no}}\}$ (Prop.~\ref{prop:domination}) \\
		Overlapping LZ77 & --- & $\le z^{\mathrm{ov}}_{77}\lceil\log_2(n+1)\rceil$ (Prop.~\ref{prop:overlap}) \\
		Single reference $R$ & --- & within factor $2$ of preset-history LZ77 (Thm.~\ref{thm:sandwich}) \\
		Computation & greedy RLZ & $O(M+n)$ with suffix automata (Thm.~\ref{thm:complexity}) \\
	\end{longtable}
	\normalsize

	\section{Full Experimental Tables}
	\label{app:full_tables}
	
	Tables~\ref{tab:full_synthetic_dna}--\ref{tab:full_genomic_1000_files} give the complete Scenario~A results of Section~\ref{sec:experiments} for each corpus, grouped by family. The last group of each table contains the optimistic bounds mentioned in Section~\ref{sec:protocol}. They estimate the dictionary-primed cost of Brotli, LZMA and PPMd by the marginal cost $\mathrm{len}(f(D\,x))-\mathrm{len}(f(D))$ plus the size of an empty stream, where $D$ is the dictionary and $x$ a record. They are not measurements: the entropy tables of the compressor are already adapted to $D$ when $x$ is appended, whereas an independently decoded record must carry or relearn them. We validated the estimator on zlib-9 and zstd-19, whose dictionary interfaces give the true cost (one draw, $2$\,KiB sampled dictionary): it under-estimated the payload by $10\%$ and $33\%$ on the genomic generator and by $36\%$ and $46\%$ on the JSON generator, respectively. Rows of this group must therefore not be compared with the others.

	\begingroup
\footnotesize
\setlength{\tabcolsep}{3.5pt}
\begin{longtable}{lrrrrrr}
\caption{Scenario~A, full results on the Synth. genomic corpus (mean record length $\approx288$\,B). Payload: sum of compressed record sizes; Index: one variable-length integer per record; Dict.\ (raw) / (stored): dictionary, codec header or model, uncompressed / as stored (zlib); Total $=$ payload $+$ index $+$ Dict.\ (raw); $\Delta=1-\text{codec}/\text{method}$ (positive: codec smaller), mean of the per-draw gains. Every method is charged for its dictionary, header or model at its \emph{raw} (uncompressed) size: for the codec, the whole uncompressed header (alphabet, dictionary $S$ and probability tables); for the baselines, the raw dictionary text; for the context controls, the raw serialized model.}
\label{tab:full_synthetic_dna} \\
\toprule
Method & Payload & Index & Dict.\ (raw) & Dict.\ (stored) & Total (B) & $\Delta$ (\%) \\
\midrule
\endfirsthead
\multicolumn{7}{c}{{\tablename\ \thetable{} -- suite de la page précédente}} \\
\toprule
Method & Payload & Index & Dict.\ (raw) & Dict.\ (stored) & Total (B) & $\Delta$ (\%) \\
\midrule
\endhead
\midrule
\multicolumn{7}{r}{\emph{Suite sur la page suivante $\rightarrow$}} \\
\endfoot
\bottomrule
\endlastfoot
\multicolumn{7}{l}{\emph{Codec}} \\
\textbf{Codec} (MDL, exact pruning) & 8\,964 & 979 & 904 & 288 & 10\,847 & ref. \\
Codec, lazy pruning ($m{=}1$) & 8\,697 & 979 & 1\,628 & 354 & 11\,304 & +4.0 \\
\midrule
\multicolumn{7}{l}{\emph{Codec ablations}} \\
Codec, $S$ = chunks of zstd-trained dict. & 9\,727 & 979 & 916 & 441 & 11\,623 & +6.7 \\
Codec w/o dictionary ($S{=}\emptyset$) & 22\,155 & 979 & 129 & 132 & 23\,263 & +53.4 \\
Codec w/o self-reference (RLZ) & 58\,391 & 1\,068 & 896 & 302 & 60\,355 & +82.0 \\
\midrule
\multicolumn{7}{l}{\emph{Dictionary-based baselines}} \\
zlib-9 + MDL dict.\ $S$ & 9\,183 & 979 & 683 & 87 & 10\,844 & 0.0 \\
zlib-9 + trained dict. & 9\,230 & 979 & 668 & 196 & 10\,877 & +0.3 \\
zlib-9 + sampled dict. & 10\,244 & 979 & 677 & 55 & 11\,901 & +8.8 \\
zstd-19 + trained dict. & 17\,945 & 979 & 830 & 203 & 19\,754 & +45.1 \\
zstd-19 + MDL dict.\ $S$ & 18\,583 & 979 & 683 & 87 & 20\,244 & +46.4 \\
zstd-19 + sampled dict. & 19\,402 & 979 & 357 & 47 & 20\,738 & +47.7 \\
\midrule
\multicolumn{7}{l}{\emph{General-purpose baselines (no dictionary)}} \\
Brotli-11 & 31\,149 & 979 & 0 & 0 & 32\,128 & +66.2 \\
zstd-19 & 35\,472 & 979 & 0 & 0 & 36\,451 & +70.2 \\
zlib-9 (raw deflate) & 40\,813 & 979 & 0 & 0 & 41\,792 & +74.0 \\
zlib-6 (framed) & 46\,687 & 979 & 0 & 0 & 47\,666 & +77.2 \\
LZMA (raw) & 47\,730 & 979 & 0 & 0 & 48\,709 & +77.7 \\
bzip2 & 66\,514 & 979 & 0 & 0 & 67\,493 & +83.9 \\
\midrule
\multicolumn{7}{l}{\emph{Idealized context-model controls}} \\
Order-3 context model (ideal) & 68\,349 & 1\,092 & 1\,274 & 913 & 70\,715 & +84.7 \\
Order-1 context model (ideal) & 70\,239 & 1\,062 & 81 & 81 & 71\,383 & +84.8 \\
Order-2 context model (ideal) & 70\,208 & 1\,070 & 349 & 308 & 71\,627 & +84.8 \\
\midrule
\multicolumn{7}{l}{\emph{Optimistic bounds (not measurements; see text)}} \\
Brotli-11, primed (sampled dict.) & 6\,030 & 979 & 958 & 61 & 7\,966 & $-$37.5 \\
Brotli-11, primed (MDL dict.\ $S$) & 8\,036 & 979 & 683 & 87 & 9\,698 & $-$12.6 \\
PPMd (order 8), primed (sampled dict.) & 10\,124 & 979 & 1\,338 & 70 & 12\,442 & +12.8 \\
PPMd (order 8), primed (MDL dict.\ $S$) & 11\,126 & 979 & 683 & 87 & 12\,787 & +15.2 \\
LZMA, primed (MDL dict.\ $S$) & 16\,474 & 979 & 683 & 87 & 18\,136 & +40.2 \\
LZMA, primed (sampled dict.) & 17\,937 & 979 & 357 & 47 & 19\,273 & +43.7 \\
\end{longtable}
\endgroup
	\begingroup
\footnotesize
\setlength{\tabcolsep}{3.5pt}
\begin{longtable}{lrrrrrr}
\caption{Scenario~A, full results on the Synth. JSON corpus (mean record length $\approx114$\,B). Payload: sum of compressed record sizes; Index: one variable-length integer per record; Dict.\ (raw) / (stored): dictionary, codec header or model, uncompressed / as stored (zlib); Total $=$ payload $+$ index $+$ Dict.\ (raw); $\Delta=1-\text{codec}/\text{method}$ (positive: codec smaller), mean of the per-draw gains. Every method is charged for its dictionary, header or model at its \emph{raw} (uncompressed) size: for the codec, the whole uncompressed header (alphabet, dictionary $S$ and probability tables); for the baselines, the raw dictionary text; for the context controls, the raw serialized model.}
\label{tab:full_synthetic_json} \\
\toprule
Method & Payload & Index & Dict.\ (raw) & Dict.\ (stored) & Total (B) & $\Delta$ (\%) \\
\midrule
\endfirsthead
\multicolumn{7}{c}{{\tablename\ \thetable{} -- suite de la page précédente}} \\
\toprule
Method & Payload & Index & Dict.\ (raw) & Dict.\ (stored) & Total (B) & $\Delta$ (\%) \\
\midrule
\endhead
\midrule
\multicolumn{7}{r}{\emph{Suite sur la page suivante $\rightarrow$}} \\
\endfoot
\bottomrule
\endlastfoot
\multicolumn{7}{l}{\emph{Codec}} \\
\textbf{Codec} (MDL, exact pruning) & 14\,547 & 1\,000 & 2\,019 & 856 & 17\,566 & ref. \\
Codec, lazy pruning ($m{=}1$) & 14\,706 & 1\,000 & 2\,838 & 1\,120 & 18\,544 & +5.3 \\
\midrule
\multicolumn{7}{l}{\emph{Codec ablations}} \\
Codec w/o self-reference (RLZ) & 14\,551 & 1\,000 & 2\,014 & 849 & 17\,565 & 0.0 \\
Codec, $S$ = chunks of zstd-trained dict. & 15\,366 & 1\,000 & 2\,091 & 867 & 18\,456 & +4.8 \\
Codec w/o dictionary ($S{=}\emptyset$) & 72\,852 & 1\,000 & 270 & 213 & 74\,122 & +76.3 \\
\midrule
\multicolumn{7}{l}{\emph{Dictionary-based baselines}} \\
zlib-9 + sampled dict. & 19\,589 & 1\,000 & 1\,747 & 356 & 22\,336 & +21.4 \\
zlib-9 + trained dict. & 19\,651 & 1\,000 & 1\,870 & 494 & 22\,521 & +22.0 \\
zlib-9 + MDL dict.\ $S$ & 20\,439 & 1\,000 & 1\,443 & 430 & 22\,882 & +23.2 \\
zstd-19 + sampled dict. & 29\,083 & 1\,000 & 1\,747 & 356 & 31\,830 & +44.8 \\
zstd-19 + trained dict. & 28\,950 & 1\,000 & 1\,882 & 498 & 31\,831 & +44.8 \\
zstd-19 + MDL dict.\ $S$ & 30\,560 & 1\,000 & 1\,443 & 430 & 33\,003 & +46.8 \\
\midrule
\multicolumn{7}{l}{\emph{General-purpose baselines (no dictionary)}} \\
Brotli-11 & 104\,746 & 1\,000 & 0 & 0 & 105\,746 & +83.4 \\
zlib-9 (raw deflate) & 107\,863 & 1\,000 & 0 & 0 & 108\,863 & +83.9 \\
LZMA (raw) & 109\,163 & 1\,000 & 0 & 0 & 110\,163 & +84.0 \\
zstd-19 & 110\,007 & 1\,000 & 0 & 0 & 111\,007 & +84.2 \\
zlib-6 (framed) & 113\,863 & 1\,000 & 0 & 0 & 114\,863 & +84.7 \\
bzip2 & 135\,482 & 2\,000 & 0 & 0 & 137\,482 & +87.2 \\
\midrule
\multicolumn{7}{l}{\emph{Idealized context-model controls}} \\
Order-2 context model (ideal) & 13\,128 & 1\,000 & 4\,186 & 2\,542 & 18\,314 & +4.1 \\
Order-3 context model (ideal) & 9\,943 & 1\,000 & 8\,595 & 4\,701 & 19\,538 & +10.1 \\
Order-1 context model (ideal) & 28\,511 & 1\,000 & 878 & 744 & 30\,389 & +42.2 \\
\midrule
\multicolumn{7}{l}{\emph{Optimistic bounds (not measurements; see text)}} \\
Brotli-11, primed (sampled dict.) & 3\,482 & 1\,000 & 3\,841 & 567 & 8\,323 & +277.3 \\
PPMd (order 8), primed (sampled dict.) & 13\,780 & 1\,000 & 1\,974 & 380 & 16\,754 & $-$4.9 \\
PPMd (order 8), primed (MDL dict.\ $S$) & 15\,706 & 1\,000 & 1\,443 & 430 & 18\,149 & +3.1 \\
Brotli-11, primed (MDL dict.\ $S$) & 15\,784 & 1\,000 & 1\,443 & 430 & 18\,227 & +2.7 \\
LZMA, primed (sampled dict.) & 20\,979 & 1\,000 & 1\,974 & 380 & 23\,953 & +26.6 \\
LZMA, primed (MDL dict.\ $S$) & 26\,110 & 1\,000 & 1\,443 & 430 & 28\,553 & +38.5 \\
\end{longtable}
\endgroup
	\begingroup
\footnotesize
\setlength{\tabcolsep}{3.5pt}
\begin{longtable}{lrrrrrr}
\caption{Scenario~A, full results on the Linux logs corpus (mean record length $\approx191$\,B). Payload: sum of compressed record sizes; Index: one variable-length integer per record; Dict.\ (raw) / (stored): dictionary, codec header or model, uncompressed / as stored (zlib); Total $=$ payload $+$ index $+$ Dict.\ (raw); $\Delta=1-\text{codec}/\text{method}$ (positive: codec smaller), mean of the per-draw gains. Every method is charged for its dictionary, header or model at its \emph{raw} (uncompressed) size: for the codec, the whole uncompressed header (alphabet, dictionary $S$ and probability tables); for the baselines, the raw dictionary text; for the context controls, the raw serialized model.}
\label{tab:full_Linux_2k_log_structured} \\
\toprule
Method & Payload & Index & Dict.\ (raw) & Dict.\ (stored) & Total (B) & $\Delta$ (\%) \\
\midrule
\endfirsthead
\multicolumn{7}{c}{{\tablename\ \thetable{} -- suite de la page précédente}} \\
\toprule
Method & Payload & Index & Dict.\ (raw) & Dict.\ (stored) & Total (B) & $\Delta$ (\%) \\
\midrule
\endhead
\midrule
\multicolumn{7}{r}{\emph{Suite sur la page suivante $\rightarrow$}} \\
\endfoot
\bottomrule
\endlastfoot
\multicolumn{7}{l}{\emph{Codec}} \\
\textbf{Codec} (MDL, exact pruning) & 25\,181 & 1\,000 & 5\,934 & 2\,244 & 32\,115 & ref. \\
Codec, lazy pruning ($m{=}1$) & 24\,497 & 1\,000 & 6\,758 & 2\,524 & 32\,255 & +0.4 \\
\midrule
\multicolumn{7}{l}{\emph{Codec ablations}} \\
Codec w/o self-reference (RLZ) & 25\,404 & 1\,000 & 5\,901 & 2\,210 & 32\,305 & +0.6 \\
Codec, $S$ = chunks of zstd-trained dict. & 31\,056 & 1\,000 & 6\,484 & 2\,463 & 38\,539 & +16.7 \\
Codec w/o dictionary ($S{=}\emptyset$) & 110\,520 & 1\,282 & 423 & 357 & 112\,225 & +71.4 \\
\midrule
\multicolumn{7}{l}{\emph{Dictionary-based baselines}} \\
zlib-9 + MDL dict.\ $S$ & 30\,285 & 1\,000 & 5\,109 & 1\,623 & 36\,393 & +11.8 \\
zlib-9 + trained dict. & 29\,501 & 1\,000 & 6\,669 & 1\,811 & 37\,170 & +13.6 \\
zlib-9 + sampled dict. & 32\,010 & 1\,000 & 7\,475 & 1\,446 & 40\,485 & +20.5 \\
zstd-19 + trained dict. & 37\,259 & 1\,000 & 7\,142 & 1\,878 & 45\,402 & +29.3 \\
zstd-19 + MDL dict.\ $S$ & 41\,663 & 1\,000 & 5\,109 & 1\,623 & 47\,772 & +32.8 \\
zstd-19 + sampled dict. & 42\,416 & 1\,000 & 7\,664 & 1\,472 & 51\,081 & +37.0 \\
\midrule
\multicolumn{7}{l}{\emph{General-purpose baselines (no dictionary)}} \\
Brotli-11 & 144\,367 & 1\,610 & 0 & 0 & 145\,976 & +78.0 \\
zlib-9 (raw deflate) & 146\,860 & 1\,793 & 0 & 0 & 148\,653 & +78.4 \\
zstd-19 & 152\,244 & 1\,938 & 0 & 0 & 154\,182 & +79.2 \\
zlib-6 (framed) & 152\,860 & 1\,925 & 0 & 0 & 154\,785 & +79.2 \\
LZMA (raw) & 153\,302 & 1\,885 & 0 & 0 & 155\,187 & +79.3 \\
bzip2 & 181\,323 & 2\,000 & 0 & 0 & 183\,323 & +82.5 \\
\midrule
\multicolumn{7}{l}{\emph{Idealized context-model controls}} \\
Order-2 context model (ideal) & 33\,099 & 1\,000 & 8\,170 & 5\,755 & 42\,269 & +24.0 \\
Order-3 context model (ideal) & 22\,965 & 1\,000 & 18\,863 & 12\,119 & 42\,829 & +25.0 \\
Order-1 context model (ideal) & 64\,131 & 1\,000 & 1\,921 & 1\,620 & 67\,052 & +52.1 \\
\midrule
\multicolumn{7}{l}{\emph{Optimistic bounds (not measurements; see text)}} \\
Brotli-11, primed (sampled dict.) & 21\,662 & 1\,000 & 5\,312 & 1\,146 & 27\,974 & $-$15.4 \\
PPMd (order 8), primed (MDL dict.\ $S$) & 23\,339 & 1\,000 & 5\,109 & 1\,623 & 29\,448 & $-$9.1 \\
Brotli-11, primed (MDL dict.\ $S$) & 24\,050 & 1\,000 & 5\,109 & 1\,623 & 30\,158 & $-$6.6 \\
PPMd (order 8), primed (sampled dict.) & 25\,234 & 1\,000 & 5\,990 & 1\,247 & 32\,224 & +0.2 \\
LZMA, primed (MDL dict.\ $S$) & 32\,404 & 1\,000 & 5\,109 & 1\,623 & 38\,513 & +16.6 \\
LZMA, primed (sampled dict.) & 31\,202 & 1\,000 & 6\,329 & 1\,289 & 38\,532 & +16.5 \\
\end{longtable}
\endgroup
	\begingroup
\footnotesize
\setlength{\tabcolsep}{3.5pt}
\begin{longtable}{lrrrrrr}
\caption{Scenario~A, full results on the Real genomic corpus (mean record length $\approx200$\,B). Payload: sum of compressed record sizes; Index: one variable-length integer per record; Dict.\ (raw) / (stored): dictionary, codec header or model, uncompressed / as stored (zlib); Total $=$ payload $+$ index $+$ Dict.\ (raw); $\Delta=1-\text{codec}/\text{method}$ (positive: codec smaller), mean of the per-draw gains. Every method is charged for its dictionary, header or model at its \emph{raw} (uncompressed) size: for the codec, the whole uncompressed header (alphabet, dictionary $S$ and probability tables); for the baselines, the raw dictionary text; for the context controls, the raw serialized model.}
\label{tab:full_genomic_1000_files} \\
\toprule
Method & Payload & Index & Dict.\ (raw) & Dict.\ (stored) & Total (B) & $\Delta$ (\%) \\
\midrule
\endfirsthead
\multicolumn{7}{c}{{\tablename\ \thetable{} -- suite de la page précédente}} \\
\toprule
Method & Payload & Index & Dict.\ (raw) & Dict.\ (stored) & Total (B) & $\Delta$ (\%) \\
\midrule
\endhead
\midrule
\multicolumn{7}{r}{\emph{Suite sur la page suivante $\rightarrow$}} \\
\endfoot
\bottomrule
\endlastfoot
\multicolumn{7}{l}{\emph{Codec}} \\
\textbf{Codec} (MDL, exact pruning) & 57\,797 & 771 & 9\,495 & 5\,148 & 68\,063 & ref. \\
Codec, lazy pruning ($m{=}1$) & 58\,069 & 771 & 15\,816 & 8\,134 & 74\,656 & +8.8 \\
\midrule
\multicolumn{7}{l}{\emph{Codec ablations}} \\
Codec, $S$ = chunks of zstd-trained dict. & 54\,959 & 771 & 7\,468 & 3\,693 & 63\,198 & $-$7.7 \\
Codec w/o self-reference (RLZ) & 56\,051 & 772 & 9\,334 & 5\,036 & 66\,157 & $-$2.9 \\
Codec w/o dictionary ($S{=}\emptyset$) & 81\,321 & 776 & 544 & 424 & 82\,641 & +17.6 \\
\midrule
\multicolumn{7}{l}{\emph{Dictionary-based baselines}} \\
zstd-19 + trained dict. & 65\,175 & 771 & 7\,348 & 3\,222 & 73\,295 & +7.1 \\
zlib-9 + trained dict. & 66\,002 & 772 & 7\,101 & 3\,123 & 73\,874 & +7.9 \\
zlib-9 + MDL dict.\ $S$ & 67\,256 & 772 & 6\,306 & 2\,742 & 74\,333 & +8.4 \\
zlib-9 + sampled dict. & 66\,543 & 772 & 7\,520 & 2\,922 & 74\,835 & +9.0 \\
zstd-19 + sampled dict. & 78\,120 & 778 & 7\,240 & 2\,824 & 86\,139 & +21.0 \\
zstd-19 + MDL dict.\ $S$ & 79\,817 & 780 & 6\,306 & 2\,742 & 86\,903 & +21.7 \\
\midrule
\multicolumn{7}{l}{\emph{General-purpose baselines (no dictionary)}} \\
Brotli-11 & 102\,545 & 1\,313 & 0 & 0 & 103\,859 & +34.5 \\
zlib-9 (raw deflate) & 108\,673 & 1\,526 & 0 & 0 & 110\,199 & +38.2 \\
zstd-19 & 109\,109 & 1\,484 & 0 & 0 & 110\,594 & +38.5 \\
zlib-6 (framed) & 113\,294 & 1\,538 & 0 & 0 & 114\,832 & +40.7 \\
LZMA (raw) & 114\,037 & 1\,534 & 0 & 0 & 115\,571 & +41.1 \\
bzip2 & 132\,221 & 1\,539 & 0 & 0 & 133\,760 & +49.1 \\
\midrule
\multicolumn{7}{l}{\emph{Idealized context-model controls}} \\
Order-2 context model (ideal) & 43\,562 & 771 & 9\,743 & 6\,665 & 54\,077 & $-$25.9 \\
Order-1 context model (ideal) & 51\,807 & 771 & 2\,550 & 1\,997 & 55\,129 & $-$23.5 \\
Order-3 context model (ideal) & 41\,004 & 771 & 18\,829 & 12\,222 & 60\,604 & $-$12.3 \\
\midrule
\multicolumn{7}{l}{\emph{Optimistic bounds (not measurements; see text)}} \\
Brotli-11, primed (sampled dict.) & 29\,784 & 770 & 7\,400 & 2\,867 & 37\,954 & $-$152.7 \\
PPMd (order 8), primed (MDL dict.\ $S$) & 47\,533 & 770 & 6\,306 & 2\,742 & 54\,609 & $-$24.6 \\
PPMd (order 8), primed (sampled dict.) & 48\,779 & 770 & 5\,360 & 2\,150 & 54\,909 & $-$24.0 \\
Brotli-11, primed (MDL dict.\ $S$) & 50\,964 & 771 & 6\,306 & 2\,742 & 58\,041 & $-$18.2 \\
LZMA, primed (sampled dict.) & 57\,331 & 770 & 6\,520 & 2\,571 & 64\,621 & $-$5.3 \\
LZMA, primed (MDL dict.\ $S$) & 58\,807 & 770 & 6\,306 & 2\,742 & 65\,883 & $-$3.4 \\
\end{longtable}
\endgroup

\bibliographystyle{IEEEtran}
\bibliography{biblio_TIT_consolidated}

\end{document}